\documentclass{article}

\usepackage{arxiv}

\usepackage{hyperref}
\hypersetup{
    colorlinks=true,
    citecolor=black,
    linkcolor=black,
    urlcolor=blue
}
\usepackage[round]{natbib}
\usepackage[utf8]{inputenc} % allow utf-8 input
\usepackage[T1]{fontenc}    % use 8-bit T1 fonts
\usepackage{url}            % simple URL typesetting
\usepackage{booktabs}       % professional-quality tables
\usepackage{amsfonts}       % blackboard math symbols
\usepackage{nicefrac}       % compact symbols for 1/2, etc.
\usepackage{microtype}      % microtypography
\usepackage{lipsum}		% Can be removed after putting your text content
\usepackage{graphicx}
\usepackage{doi}
\usepackage{amsmath}
\usepackage{longtable}
\usepackage{multirow}
\usepackage{multicol}
\usepackage{subcaption}
\usepackage{array}
\usepackage{placeins}

\title{Markov Decision Process Approximation Methods for Water Distribution Network Inspection and Maintenance: A Case Study of the U.S.~Virgin Islands}

\author{Minsuk Seo\\
	Republic of Korea Army\\
	South Korea\\
    \texttt{smith44189@gmail.com} \\
	\And
	Daniel Eisenberg \\
	Department of Operations Research\\
	Naval Postgraduate School\\
	Monterey, CA, USA \\
	\texttt{daniel.eisenberg@nps.edu} \\
    \And
	Jefferson Huang \\
	Department of Operations Research\\
	Naval Postgraduate School\\
	Monterey, CA, USA \\
	\texttt{jefferson.huang@nps.edu} \\
}

\renewcommand{\headeright}{Seo et al., 2026}
\renewcommand{\undertitle}{}
\renewcommand{\shorttitle}{MDP Approximation for Water Network Maintenance}

\hypersetup{
pdftitle={MDP Approximation for Water Network Maintenance},
pdfsubject={q-bio.NC, q-bio.QM},
pdfauthor={Seo et al.},
pdfkeywords={Water Distribution Network, Markov Decision Process, Optimal Maintenance Policy, Water Network Tool for Resilience, U.S.~Virgin Islands},
}

\begin{document}
\maketitle

\begin{abstract}
We develop a repair-oriented inspection and maintenance decision framework for water distribution networks. This work is motivated by utilities operating in data-sparse environments, such as in remote locations like the U.S.~Virgin Islands, where data collection about network state and underground pipeline outages is limited to above-ground and easy to access information (e.g., water tank levels and pump operations). We formulate the problem as a discounted Markov decision process and integrate it with high-fidelity hydraulic simulation. The model captures latent system dynamics without requiring pipe-level sensing. The results reveal state-dependent optimal policies and heterogeneous failure characteristics across pipes, including rare but high-impact behaviors. We further show that certain observable system states uniquely correspond to specific pipe failures, enabling a form of virtual sensing. These findings demonstrate that system-level dynamics can support inspection planning and maintenance decisions under uncertainty in resource-constrained settings.
\end{abstract}

% keywords can be removed

%%%%%%%%%%%%%%%%%%%%%%%%%%%%%%%%%%%%%%%
%%%%%%%%%%%%%%%%%%%%%%%%%%%%%%%%%%%%%%%
%%%%%%%%%%%%%%%%%%%%%%%%%%%%%%%%%%%%%%%
% Main text
\section{Introduction}\label{sec:intro}

Formulating effective maintenance decisions for critical infrastructure networks presents a formidable challenge for utility service providers \citep{eisenberg_rethinking_2019}. For the purposes of this work, maintenance decisions refer to repair actions meant to keep aging and failed assets functioning. In conventional management systems, decision-making typically relies on factors evaluated independently for each asset, such as condition, degradation pattern, failure rates, and resource availability \citep{andersen_numerical_2022, frangopol_probabilistic_2004, bhardwaj_energy_2023}. Optimal maintenance decisions in this setting are only constrained by resources (e.g., time, cost, equipment, and manpower).

When the assets are connected into networks to deliver critical services (e.g., pipelines and power grids), repair decisions can have far greater complexity due to the high interdependence of components. Factors such as time delayed impacts, uncertain outage costs, and latent resource constraints complicate analysis and optimization of maintenance policies \citep{alderson_assessing_2014, dickenson_interdependent_2014, thomas_mixedinteger_2024}. For example, in a water distribution network (WDN), cascading failures can propagate with time delays, such that certain segments of the network may not immediately experience the effects of a localized disruption. This means large leaks or pipe failures can exist in a WDN, yet customers may not experience low pressures or outages, even if they are imminent \citep{bunn_cascading_2018}. Failure to respond prior to a cascade can cause widespread outages from only a few component failures \citep{shuang_node_2014}. Optimal maintenance decisions must be able to identify and remediate these highly vulnerable operating states \citep{alderson_operational_2015}.

Maintenance decisions are further complicated for utility providers in remote environments with limited equipment or manpower for data collection on the system state \citep{borgdorff_measuring_2020}. Many critical infrastructure network assets are buried underground, which can make it difficult to assess their operating condition \citep{alderson_interdependent_2018, gas_pomdp_2020}. Utilities in remote regions also cannot afford to install, maintain, and analyze complex system control and data acquisition (SCADA) systems, and must rely on basic operational data \citep{USVI_water_report_2022, alderson_interdependent_2018,
caribbean_report_2024}. For a WDN, this means customer demand, flow meters, valves, or otherwise in-line asset conditions may not be available. Instead, maintenance data are limited to easily accessible information, such as repair history, water tank levels, pump operations, and customer outage calls.

Infrastructure resilience studies establish the importance of maintaining critical services under disruption, emphasizing operational performance \citep{alderson_operational_2015}, interdependencies \citep{alderson_assessing_2014}, and recovery dynamics \citep{brendecke_optimal_2016}. These studies provide valuable insights into system vulnerability and resilience metrics, particularly for interdependent infrastructure systems such as those in the U.S.~Virgin Islands (USVI) \citep{alderson_interdependent_2018, klise_resilience_2022, borgdorff_measuring_2020, good_operational_2019}. However, resilience analyses are typically evaluative rather than prescriptive, focusing on scenario-based performance assessment rather than optimal maintenance or repair policies.

The objective of this work is to derive optimal maintenance policies and characterize system-level behavior in a WDN based on readily available data for utilities like those in the USVI. Accordingly, optimal maintenance decisions must be able to manage time delays, cascading effects, and outage costs experienced in WDNs.

\section{Background}\label{sec:background}

\subsection{Water Distribution Networks}
A WDN consists of interconnected components---such as pipes, tanks, valves, pumps, reservoirs, and junctions---that ensure the reliable delivery of potable water while maintaining hydraulic and operational constraints \citep{rossman_epanet_2000}. The dynamics of a WDN are constrained by physical laws, including mass conservation, energy balance, and pressure-flow relationships, which establish how water flows through the network and how service availability is affected by component failures or operational disruptions. Analyzing WDNs entails significant computational complexity due to inherent non-linearity in hydraulic relationships. In particular, nonlinear pressure losses and energy interactions across pipes and pumps make the resulting system difficult to analyze \citep{wang_geometric_2019}. 

Hydraulic simulation models are widely used in the water industry to analyze the function of a WDN. Currently, the most common open-source tool for hydraulic simulation is the Water Network Tool for Resilience (WNTR) \citep{Klise2018WNTR}, which enables both hydraulic simulation and resilience analysis of WDNs. WNTR is built upon the Environmental Protection Agency Network Evaluation Tool (EPANET), which is an application for modeling WDNs that was developed by the Environmental Protection Agency (EPA). Hydraulic simulation resolves non-linear losses and flows via hydraulic balancing and user-defined operational controls (e.g., pump and tank settings). A key metric of system performance is water service availability (WSA), which is defined as the ratio between the water delivered to a demand node (i.e., junction) and the expected demand \citep{klise2020wntr}. This metric captures service disruptions caused by factors such as disconnections, component failures, and storage depletion. Hence, WNTR and related simulation models can inform the potential impacts of asset outages due to lack of maintenance and failure. 

\subsection{Optimal Maintenance of Water Infrastructure and Networks}

To determine optimal maintenance plans, two methodological approaches have emerged: condition-based maintenance (CBM) and simulation-based water network optimization. CBM has developed substantial decision-theoretic frameworks using Markov decision processes (MDPs) and partially observable Markov decision processes (POMDPs) \citep{nguyen_joint_2019} to model deterioration \citep{zhang_geometric_2002}, inspection \citep{osman_optimizing_2012}, and maintenance \citep{andersen_numerical_2022} actions. These models offer mathematically principled policies and have been extended using reinforcement learning (RL) and multi-agent formulations \citep{andriotis_optimizing_2021, ferreira_neto_deep_2024, najafi_deep_2023, do_multi-agent_2024}. Nevertheless, CBM studies generally treat components independently \citep{do_multi-agent_2024, osman_optimizing_2012} or assume simplified dependency structures \citep{habib_efficient_2026}, and rarely incorporate network hydraulics in system objectives. As a result, their applicability to complex interconnected WDNs remains limited.

Simulation-based water network optimization has advanced significantly in terms of computational efficiency and physical realism \citep{wang_new_2020, guo_optimal_2023}. Approaches based on robust optimization, approximation techniques \citep{tasseff_polyhedral_2024, ulusoy_distributed_2025}, and hydraulic simulation tools such as EPANET and WNTR enable the analysis of large-scale, realistic networks \citep{thomas_magnets_2023, thomas_mixedinteger_2024}. Despite these strengths, most simulation-based
studies focus on short-term operational control, such as pump or valve scheduling, and do not yield explicit long-term maintenance or repair policies \citep{guo_optimal_2023, ulusoy_distributed_2025}. Moreover, the lack of a formal state-based decision framework makes it difficult to prescribe proactive actions under uncertainty.

Recent studies applying MDPs and RL to water infrastructure represent an important step toward unifying decision-theoretic and hydraulic perspectives \citep{restelli_optimal_2022, salaorni2021optimal, hu_multi-objective_2022}. However, existing work mostly addresses operational control problems \citep{donancio_pump_2025, battiti_intelligent_2019, guo_optimal_2023, ulusoy_distributed_2025} or applies maintenance decisions to simplified or synthetic networks \citep{stisser_comparison_2025, hajgato_deep_2020}. In addition, most studies assume Markovian dynamics without empirical validation, and state definitions are often selected in an ad-hoc manner, limiting interpretability and policy transferability.

In summary, there exists a gap at the intersection of these research streams. No prior work has developed a policy-oriented MDP framework that:

\begin{enumerate}
    \item explicitly integrates repair-oriented maintenance decisions with high-fidelity, physics-based hydraulic simulation;
    \item validates Markov assumptions through empirical testing; and,
    \item incorporates mission-oriented objectives that reflect service availability and operational resilience.
\end{enumerate}

This research addresses these gaps by formulating a discounted MDP for repair decisions in a potable WDN in the USVI. By integrating WNTR-based hydraulic simulation with statistically validated state definitions and operation-oriented cost structures, the proposed framework bridges CBM modeling and simulation-based water network optimization. Together, it provides actionable policy-level decision support for infrastructure resilience and identifies how decision policies change with changing key input variables (i.e., pipe failure probability and repair cost).

\section{Problem Statement}\label{sec:problem}

We develop a method to optimize maintenance decision policies in water infrastructure systems, focusing on repair decisions while supporting inspection planning. Our model objective is to minimize life-cycle maintenance costs while maintaining critical service availability. We structure this problem as an MDP---a sequential decision-making process in which the decision-maker must determine when to inspect system components and when to perform repair actions under uncertainty. We apply our methods to study the potable WDN connecting the islands of St.~Thomas (STT) and St.~John (STJ) in the USVI---an isolated system with limited inspection and repair resources that has similar features to military installations or forward operating bases in austere environments.

\begin{figure}[!ht]
    \centering
    \includegraphics[width=.5\linewidth]{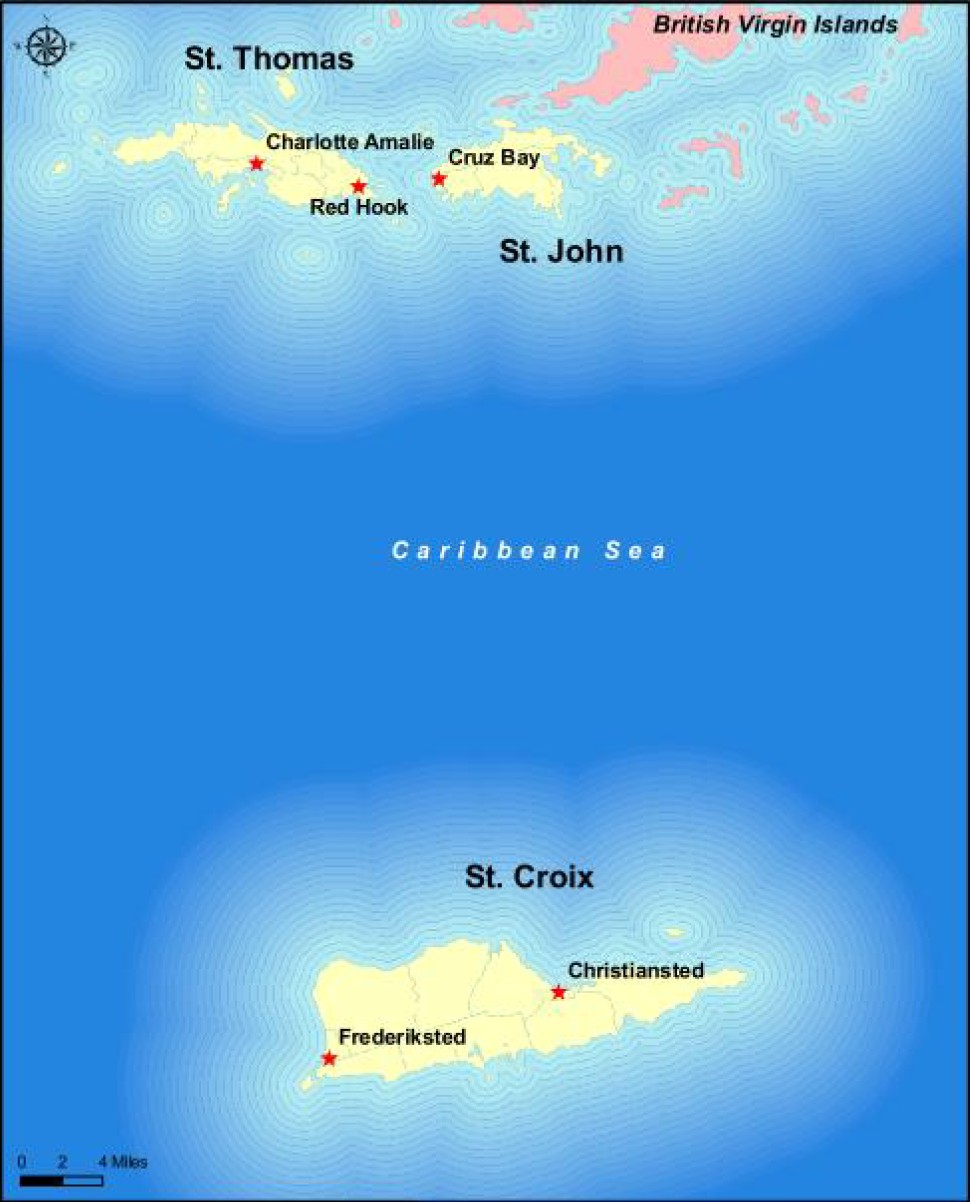}
    \caption[USVI territory]{USVI territory; The three main islands in the USVI: STT, STJ and STX. Source: \citet{usvi_transportation_2014}}
    \label{fig:USVI_Territory} 
\end{figure}

\subsection{Markov Decision Process Assumptions}

An MDP models a system evolving over discrete time steps, where decisions influence probabilistic state transitions. Under the Markov property, the system dynamics depend only on the current state and not on the past history \citep{puterman_markov_2009}. At each time step, the decision-maker observes the system state, selects an action (e.g., \textit{Repair} or \textit{DoNothing}), incurs an immediate cost, and the system transitions stochastically to a new state. The objective is to find a policy that minimizes the expected total cost over the planning horizon. Applying an MDP to WDNs is nontrivial due to specific features of WDNs. In particular, critical components such as pipelines are typically underground and may continue to operate in degraded conditions without direct observability \citep{bunn_cascading_2018, shuang_node_2014}. Moreover, due to storage in tanks and reservoirs, the impact of component failures is often delayed, introducing temporal dependencies that complicate state representation and transition modeling \citep{eisenberg_rethinking_2019, alderson_interdependent_2018}.

This study formulates the maintenance decision problem as an MDP under two standard assumptions. First, we model component failures as age-independent stochastic processes. Second, we adopt a perfect repair assumption, under which a maintenance action restores a component to an `as-good-as-new' condition \citep{frangopol_probabilistic_2004}. Under these assumptions, the decision problem reduces to determining when to perform repair actions versus when to continue operating the system in a degraded state. An MDP provides a natural framework for modeling such sequential decision-making under uncertainty, where operational and maintenance costs are incurred over time. These assumptions are simplified compared to other CBM models structured as MDPs which include degraded asset states in-between fully operational and fully failed. However, our methods can be easily extended to include these additional states in the analysis.

\subsection{Case Study Setting: The U.S.~Virgin Islands}

This study focuses on the WDN of the USVI, a geographically isolated U.S. territory in the Caribbean Sea. The USVI consists of three main islands—STT, STJ, and St. Croix (STX)—with a total land area of approximately 150 square miles \citep{usvi_transportation_2014}. Figure~\ref{fig:USVI_Territory} illustrates the geographic layout of the USVI.

The physical separation of these islands, limited redundancy, and constrained repair resources make the WDN particularly vulnerable to disruptions. In addition, in the absence of pipe-condition sensing, the operational status of critical components is not directly observable, making it difficult to determine whether a pipe is functional or has failed. This limited observability motivates the need for inspection and information acquisition in the decision-making process.

Recent natural disasters have exposed the fragility of the USVI’s critical infrastructure, leading to prolonged service disruptions and significant challenges in system recovery \citep{klise_resilience_2022, good_operational_2019, borgdorff_measuring_2020, alderson_interdependent_2018}. These events highlighted the importance of infrastructure resilience and motivated efforts by federal and local agencies to improve system robustness \citep{alderson_interdependent_2018}. In this context, the USVI provides a compelling case for studying maintenance decision-making under uncertainty. 

% In this study, we analyze the potable water distribution system using a quantitative modeling framework to support optimal maintenance and operational decisions.

\section{Methods}\label{sec:methods}

This work involves four integrated analysis steps:
\begin{enumerate}
    \item Selection of candidate WDN components for maintenance planning;
    \item Establishing the Markov property for potential state-action definitions and decisions;
    \item Constructing an associated MDP model; and,
    \item Model solution and analysis.
\end{enumerate}

The first step is informed by the decision-maker's needs and the WDN topology, meaning it varies significantly depending on the specific system configuration. Steps 2-4 involve standard modeling and optimization techniques for MDPs that are reproducible for other systems. One of the novel contributions of this work is the integration of these standard techniques that are rarely implemented for critical infrastructure networks or informed by output of hydraulic simulation models.

\subsection{Empirical Assessment of the Markov Property}

After selecting a set of candidate components, we identify system variables and a time step under which the Markov property is approximately satisfied in order to formulate the WDN as an MDP. We define the continuous variable $X_t$ as the tank level, measured relative to the base elevation of the tank, and the discrete control variable $U_t$ as the pump status, where each pump is represented as a binary variable (on/off). These variables capture the essential hydraulic behavior of the system, as tank levels reflect system storage capacity and pumps regulate flow dynamics \citep{rossman_epanet_2000}. Furthermore, these variables are operationally observable and routinely monitored in practice, unlike pump pressure or hidden pipe conditions \citep{diaz_topological_2017}. We evaluate multiple time steps ranging from 1 to 120 hours to identify an appropriate temporal discretization. If the time step is too small, strong temporal dependencies remain in the system dynamics, violating the Markov assumption. Conversely, if the time step is too large, it smooths out transient hydraulic responses and reduces the ability to distinguish meaningful transitions \citep{puterman_markov_2009}.

To assess the validity of the Markov property, we compare the predictive performance of first- and higher-order models. Specifically, we evaluate whether the incorporation of additional historical information improves the accuracy of the prediction. For example, the Markov property approximately holds when

\begin{equation} \label{eq:markov_comparison1}
P(X_{t+1}|X_t,U_t) \approx P(X_{t+1}|X_t,X_{t-1},U_t).
\end{equation}

\noindent Similarly, we compare against a third-order approximation:

\begin{equation} \label{eq:markov_comparison2}
P(X_{t+1}|X_t,U_t) \approx P(X_{t+1}|X_t,X_{t-1},X_{t-2},U_t).
\end{equation}

We implement time-series K-fold cross-validation by partitioning the dataset into $k$ sequential folds, ensuring that training data always precedes test data chronologically \citep{james_statistical_2013}. As the fold index increases, the training set is expanded in time to reflect the temporal structure of the data. To compare model performance, we estimate autoregressive models with different memory lengths (Equations~\ref{eq:regression1}--\ref{eq:regression3}) and compute the mean squared error (MSE) across folds. Estimating these models via ordinary least squares provides a consistent estimate of the linear model that minimizes the forecast MSE \citep{hamilton_time_1994}.

\begin{equation} \label{eq:regression1}
X_{t+1} = \beta_0 + \beta_1X_t + \beta_2^TU_t + \epsilon
\end{equation}

\begin{equation} \label{eq:regression2}
X_{t+1} = \beta'_0 + \beta'_1X_t + \beta'_2X_{t-1} + {\beta'_3}^TU_t + \epsilon
\end{equation}

\begin{equation} \label{eq:regression3}
X_{t+1} = \beta''_0 + \beta''_1X_t + \beta''_2X_{t-1} + \beta''_3X_{t-2} + {\beta''_4}^TU_t + \epsilon.
\end{equation}

Furthermore, we evaluate the Markov property under multiple pipe-operating scenarios of each critical pipe to account for changes in system dynamics. Since critical pipes play a dominant role in governing flow propagation in the network, their operational status can significantly alter the temporal evolution of the system. We define three distinct conditions: functional, failed, and repaired. This enables us to verify whether the Markov assumption holds consistently across different system regimes.

After averaging the fold-wise MSEs, we quantify the performance difference using two metrics: the improvement ratio (Equation~\ref{eq:improvement_ratio}) and the p-value of a paired t-test. The paired t-test assesses whether the mean of the paired fold differences deviates significantly from zero \citep{devore_probability_2008}.

\begin{equation} \label{eq:improvement_ratio}
\text{Improvement ratio} =
\frac{\text{MSE}_{1} - \text{MSE}_{2}}{\text{MSE}_{1}}
\end{equation}

If higher-order models do not provide statistically significant predictive improvement, the first-order Markov assumption is considered sufficient. This empirical validation enables the selection of an appropriate modeling framework for the subsequent formulation of MDP.

\subsection{Markov Decision Process Model}

After proving that the chosen variables and the time step satisfy the Markov property, we construct an MDP model for maintenance decisions in a WDN. An MDP model consists of 4 elements \citep{puterman_markov_2009}:

\begin{enumerate}
    \item $\mathcal{S}$: the set of possible states
    \item $\mathcal{A}$: the set of possible actions
    \item $\mathcal{P}$: the transition probabilities
    \item $\mathcal{C}$: the one-step cost function
    % \item $\gamma:$ is the discount factor
\end{enumerate}

\subsubsection{State and Action Representation}

We model the WDN state via information readily available in data-poor environments like the USVI. Specifically, we represent WDN state via tank data verifiable via human inspection. Hence, system operators can only know pipe state (operational or failed) indirectly via water storage dynamics. This abstraction captures the essential dynamics of WDNs, where localized storage buffers supply demand while upstream failures propagate gradually through storage depletion.

% and a critical connecting pipe, whose interaction governs service availability.

For each tank, we consider (i) the discretized level, categorized as operational (OP) or non-operational (NOP) based on a predefined threshold, and (ii) the change in level, categorized as increase (INC), decrease (DEC), or maintain (MAINT). These variables summarize the system’s hydraulic behavior and provide a compact representation of network conditions relevant for decision-making. The action space consists of two maintenance decisions applied to the critical pipe: \textit{DoNothing} and \textit{Repair}.

While physical inspection is not modeled as an explicit action, the chosen state variables inherently capture diagnostic signatures of system health. This allows the observable state transitions to function as a virtual sensing mechanism, providing a foundation for targeted inspection planning.

However, due to delayed hydraulic effects, the current observable state does not fully determine the future evolution of the system. Identical tank conditions may arise from different underlying physical states, particularly with respect to the time elapsed since a pipe failure. To capture this hidden temporal dependency without expanding the observable state space, we introduce a latent variable $\tau$ representing the elapsed time since failure. This enables a compact yet expressive state representation for subsequent decision modeling.

\subsubsection{Transition Model with Latent Failure Time}

The latent failure-time variable $\tau$, which captures the elapsed time since a pipe failure, conditionally governs the transition dynamics. Each realization of $\tau$ induces distinct system dynamics due to the time-dependent drainage and refill behavior of storage elements. As a result, the system evolution is conditionally Markovian given $\tau$, even though the observable state alone cannot fully dictate the process.

We construct the transition probability by applying the law of total probability over the latent variable:
\begin{equation}
P(s' \mid s, a) = \sum_{\tau \in \mathcal{T}} P(s' \mid s, a, \tau)\, P(\tau \mid s).
\end{equation}
This formulation represents a mixture over latent failure-time states, where each component corresponds to a distinct physical evolution of the system. The latent variable $\tau$ is independent of the action at the decision epoch, as actions are selected after the realization of $\tau$, resulting in a state-dependent mixture distribution.

This approach preserves the temporal dependencies inherent in the physical system while maintaining computational tractability. In addition, $\tau$ implicitly encodes the operational status of the pipe, eliminating the need for an explicit pipe-state variable and yielding a compact yet expressive transition model. Specifically, $\tau = -\infty$ indicates a functional pipe, while $\tau \ge 0$ represents a failed pipe.
In practice, we calculate both $P(\tau | s)$ and $P(s' | s, a, \tau)$ using a counting-based approach over physically admissible realizations.

\subsubsection{One-step Cost Structure}

We define the one-step cost function to capture both service degradation and maintenance expenditure within a unified framework. The total cost consists of two components: a flow cost that reflects service loss and a maintenance cost associated with \textit{Repair} actions. Because service degradation depends on the elapsed time since failure, we define the flow cost as the expected service loss over the latent variable $\tau$:
\begin{equation}
c(s,a) = \sum_{\tau \in \mathcal{T}} \mathbb{E}[C_{\text{flow}}(s,a,\tau)] P(\tau \mid s) + M(a).
\end{equation}

Furthermore, we define the flow cost $C_{\text{flow}}(s,a,\tau)$ using WSA values calculated from WNTR-based hydraulic simulations of the WDN. Specifically, we impose a penalty when the WSA falls below a predefined threshold, capturing service disruptions such as supply shortages or disconnections:
\begin{equation}
C_{\text{flow}}(s,a,\tau) = \sum_{i \in \mathcal{J}} C_i \cdot 1 \{WSA_i (s,a,\tau) \leq \theta_i \}.
\end{equation}

Hydraulic simulation is integrated into the framework through the computation of WSA values and the empirical estimation of service degradation costs using WNTR. This binary threshold-based formulation reflects the operational perspective that service degradation becomes critical only when availability drops below an acceptable level. We define threshold values empirically using hydraulic simulation results to ensure consistency with system behavior. 

The maintenance cost $M(a)$ represents the direct cost of \textit{Repair} actions. Under the perfect repair assumption, a fixed cost is incurred when the \textit{Repair} action is selected, and zero otherwise. This cost structure enables the model to capture the trade-off between immediate repair costs and future service degradation, allowing the MDP to identify optimal maintenance policies that balance operational performance and resource expenditure.

\subsubsection{Computing an Optimal Policy}

We compute the optimal maintenance policy by formulating the problem as a discounted infinite-horizon MDP that minimizes the expected cumulative cost. We introduce a discount factor $\gamma \in [0,1)$ to ensure that the objective remains finite and to balance immediate and future costs. Accordingly, we define the objective as minimizing the expected discounted sum of one-step costs over time.

We exploit the fact that a discounted MDP with finite state and action spaces admits an optimal stationary deterministic policy. We reformulate the problem as a linear program (LP) using state–action occupancy measures \citep{puterman_markov_2009}. The occupancy measure $x_{s,a}$ represents the expected discounted number of times the policy selects action $a$ in state $s$. We then minimize the total expected cost subject to flow conservation constraints that enforce consistency with the transition dynamics:

\begin{equation}
\begin{aligned}\label{eq:bellman_cost_lp}
\underset{x}{\text{minimize}} \quad &
\sum_{s \in \mathcal{S}} \sum_{a \in \mathcal{A}} c(s, a) \, x_{s,a} \\
\text{subject to} \quad &
\sum_{a \in \mathcal{A}} x_{s,a}
- \gamma \sum_{s' \in \mathcal{S}} \sum_{a' \in \mathcal{A}}
p(s \mid s', a') \, x_{s',a'} = 1, \quad \forall s \in \mathcal{S}, \\
& x_{s,a} \geq 0 \quad \forall s \in \mathcal{S}, a \in \mathcal{A}.
\end{aligned}
\end{equation}

We recover the optimal policy by normalizing the optimal occupancy measures across actions for each state. This formulation produces an exact solution, unlike sampling-based reinforcement learning methods that approximate the optimal policy. We implement the LP in Pyomo and solve it using the Gurobi optimizer based on the explicitly constructed transition model.

We define the optimal policy $\pi^*$ as a stationary policy that minimizes the expected discounted cumulative cost:

\begin{equation}
\pi^* \in \arg\min_{\pi} 
\mathbb{E}^{\pi}\!\left[
\sum_{t=0}^{\infty} \gamma^t c(s_t,a_t)
\right].
\end{equation}

\subsection{Maintenance Policy Analysis}

We analyze how optimal maintenance policies change under different model hyperparameters, including the discount factor, the relative maintenance cost weight, and the probability of pipe failure.

The discount factor $\gamma$ regulates the relative importance of future costs compared to immediate costs \citep{puterman_markov_2009}. A larger $\gamma$ places greater emphasis on long-term system performance by assigning similar importance to future and present costs. In contrast, a smaller $\gamma$ prioritizes minimizing short-term costs by heavily discounting future consequences.

Furthermore, the optimal policy depends on the trade-off between flow cost and maintenance cost. If the flow cost associated with service degradation is substantially larger than the maintenance cost, the policy is more likely to select the \textit{Repair} action across a broader set of states. Conversely, if the maintenance cost dominates, the policy tends to postpone \textit{Repair} actions and continue operating the system in degraded conditions. To analyze this trade-off, we introduce a nonnegative weight parameter $\lambda$ applied to the maintenance cost component:

\begin{equation} \label{one-step cost_sensitivity}
c(s,a) = Flow\_cost(s) + \lambda \cdot Maintenance\_cost(a).
\end{equation}

In addition, the probability of pipe failure $p_{fail}$ directly affects the transition dynamics of the MDP. The higher $p_{fail}$ increases the likelihood that the system transitions to degraded or failed states, thereby increasing the long-term risk associated with deferred maintenance. In contrast, lower values of $p_{fail}$ increase the likelihood that the system remains in operational states, making the \textit{DoNothing} action more attractive.

To better understand how the optimal policy responds to these parameters, we perform a sensitivity analysis across a few values of $p_{fail}$ while varying $\lambda$. This analysis allows us to identify how the optimal policy structure changes between repair-oriented and no-repair regimes, and how these transitions affect the total expected cost.

\section{Results}\label{sec:results}

\subsection{System Scope and Candidate Components}

\begin{figure*}[htbp]
    \centering
    \includegraphics[width=\linewidth]{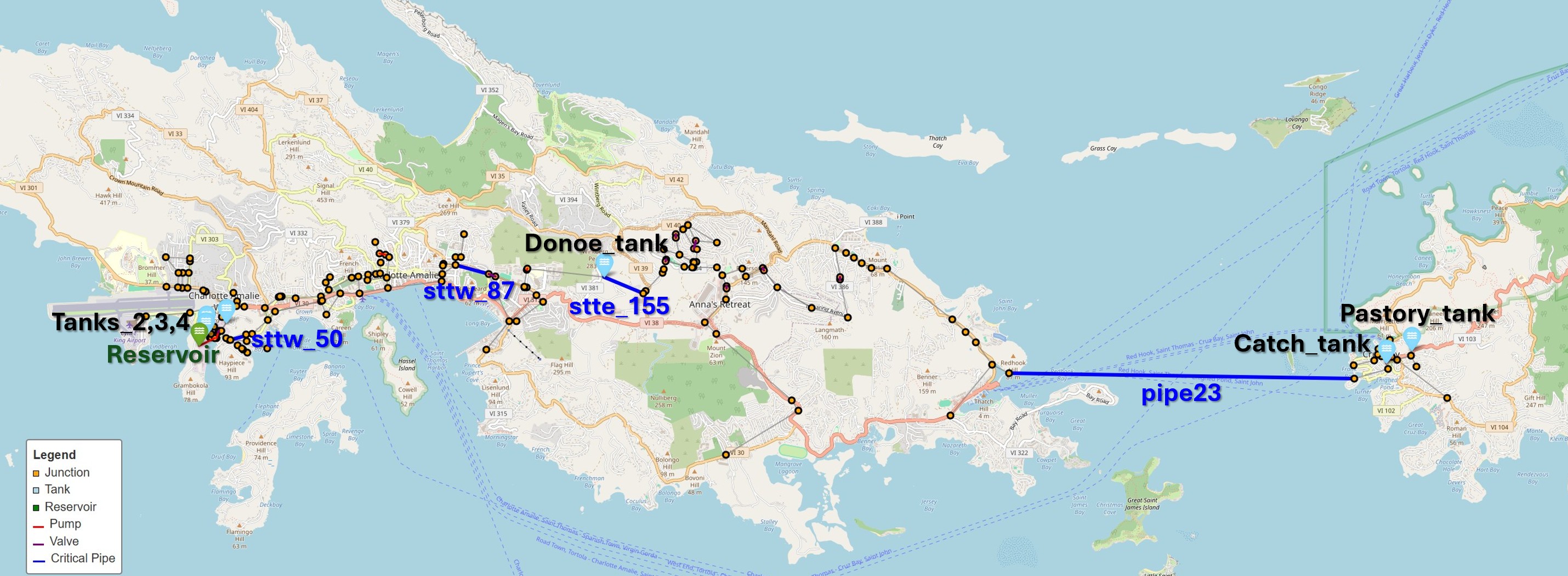}
    \caption[The St.~Thomas - St.~John water pipeline.]{The St.~Thomas - St.~John water pipeline. Water is produced in Western STT at the system reservoir via reverse osmosis and stored in Tanks 2, 3, and 4 prior to entering the pipeline system. From there, water is primarily pumped eastward to the central business district of STT (Charlotte Amalie) and Donoe tank. Water is then sent to eastern STT and STJ via underwater pipeline. Water is stored and delivered locally on STJ via Catch and Pastory tanks. Based on this system operation, four critical pipes are identified and highlighted in blue: sttw\_50, sttw\_87, stte\_155 and pipe23 (west to east). Likewise, three main storage tanks in-line with the system indicate underground and underwater pipe state: Donoe, Catch and Pastory.}
    \label{fig:STT_STJ_critical_pipes_labels} 
\end{figure*}

Figure~\ref{fig:STT_STJ_critical_pipes_labels} represents the STT-STJ WDN, which includes one reservoir, six tanks, eight pumps, eight valves, 160 junctions, and 180 pipes. Modeling all components would lead to a combinatorial explosion of state space. Therefore, we restrict the analysis to a set of candidate components selected on the basis of their influence on system dynamics and service availability. 

We first identify representative tanks by considering the functional distance from the reservoir. The reservoir is located in the far left of STT, and Tanks 2, 3, 4 are close to the reservoir and each other. This results in a relatively stable water level and limited variability from transitions, making these tanks less informative for capturing temporal dependence. In contrast, the Donoe, Catch, and Pastory tanks are located farther from the reservoir, making them more sensitive to upstream disruptions. We therefore select these three tanks as the candidate components for the Markov property tests.

Next, we identify critical pipes using both quantitative service availability (WSA) and topological characteristics. For every pipe, we simulate a failure scenario and compute the average WSA of all junctions over a 7 day period on that island. We rank pipes based on their impact on WSA and examine their topological roles. Table~\ref{tab:critical_pipes} indicates the top 5 pipes most significantly affecting the WSA of STT and STJ.

\begin{table}[!ht]
    \centering
    \caption{Top 5 most critical pipes in STT and STJ based on Average WSA}
    \label{tab:critical_pipes}
    {
    \begin{tabular}{c | l c | l c}
        \toprule
        & \multicolumn{2}{c|}{\textbf{STT Network}} & \multicolumn{2}{c}{\textbf{STJ Network}} \\
        \cmidrule{2-3} \cmidrule{4-5}
        \textbf{Rank} & \textbf{Pipe ID} & \textbf{Average WSA} & \textbf{Pipe ID} & \textbf{Average WSA} \\
        \midrule
        1 & sttw\_50 & 0.32 & pipe23 & 0.12 \\
        2 & sttw\_51 & 0.32 & stje\_28 & 0.12 \\
        3 & stte\_155 & 0.56 & stje\_1 & 0.12 \\
        4 & stte\_24 & 0.56 & stje\_15 & 0.69 \\
        5 & sttw\_87 & 0.57 & stje\_18 & 0.79 \\
        \bottomrule
    \end{tabular}%
    }
\end{table}

Of the pipes identified in Table~\ref{tab:critical_pipes}, we select four for Markov testing and our MDP model: STT west pipe 50 ($sttw\_50$), STT west pipe 87 ($sttw\_87$), STT east pipe 155 ($stte\_155$) and $pipe23$, which has a special designator between islands. These pipes are chosen for the following reasons:

\begin{enumerate}
    \item \textbf{sttw\_50} connects the reservoir and airport area (Charlotte Amalie West) to the east (Charlotte Amalie).
    \item \textbf{sttw\_87} links to a main tourist area with large demands due to cruise ships and hotels.
    \item \textbf{stte\_155} is the pipe after Donoe tank, which supplies all water to eastern STT and STJ.
    \item \textbf{pipe23} is the only underwater transmission pipe connecting STT and STJ, such that all water demand on STJ depends on it.
\end{enumerate}

\noindent Moreover, each of the other critical pipes in Table~\ref{tab:critical_pipes}, $sttw\_51$, $stte\_24$, $stje\_28$, and $stje\_1$, directly follow one of our four selected pipes for analysis. Hence, optimal repair decisions for our selected pipe segments also inform optimal decisions for the others.

% $sttw\_50$ is a topological bridge solely connecting reservoir and airport area (Charlotte Amalie West) to the east (Charlotte Amalie). $stte\_155$ is located immediately downstream of the Donoe tank, which supplies water directly to residential zones. $sttw\_87$ flows water to a main tourism area with cruises and hotels.

\subsection{Time-Step Selection via Markov Property Test}

We evaluate the validity of the Markov assumption by comparing the predictive performance of first-order and higher-order models across multiple system configurations. Specifically, we perform the analysis for each critical pipe under three operating scenarios (functional, failed, and repaired) and for three representative tanks (Donoe, Catch, and Pastory). For each configuration, we assess whether incorporating additional historical information improves prediction accuracy using both statistical and practical metrics.

Across all configurations, the results consistently indicate that higher-order models provide limited additional value beyond the first-order model at appropriate time scales. Based on this analysis, we select a 46-hour time step as a unified discretization. At this time scale, the paired t-test p-values are predominantly above the 0.05 significance threshold, indicating that the improvements from higher-order models are not statistically significant. In addition, the corresponding improvement ratios are consistently small, suggesting negligible practical gains in predictive performance.

Although a small number of configurations—particularly under pipe-failure scenarios for $stte\_155$ and $pipe23$—exhibit deviations from this pattern, these cases are limited and do not alter the overall conclusion. Importantly, the 46-hour time step appears consistently among the top-performing candidates across all pipe scenarios, providing a robust and consistent temporal resolution. These findings support the use of a first-order Markov model and justify adopting a unified time discretization for subsequent MDP formulation.

\subsection{Building the Discounted MDP}

\subsubsection{State Representation}

After selecting the MDP time step as 46 hours, we define the system state using the tank levels and their changes for the three candidate tanks:
\begin{equation}
S_t =
\bigl(
L_{t}^{D},\, \Delta L_{t}^{D},\,
L_{t}^{C},\, \Delta L_{t}^{C},\,
L_{t}^{P},\, \Delta L_{t}^{P}
\bigr),
\end{equation}
where $L_t$ denotes the tank level at decision epoch $t$, and $\Delta L_t$ denotes the change in tank level relative to the previous decision epoch.

\begin{figure}[htbp]
    \centering
    \begin{subfigure}[b]{0.48\textwidth}
        \centering        \includegraphics[width=\linewidth, height = 3.5cm]{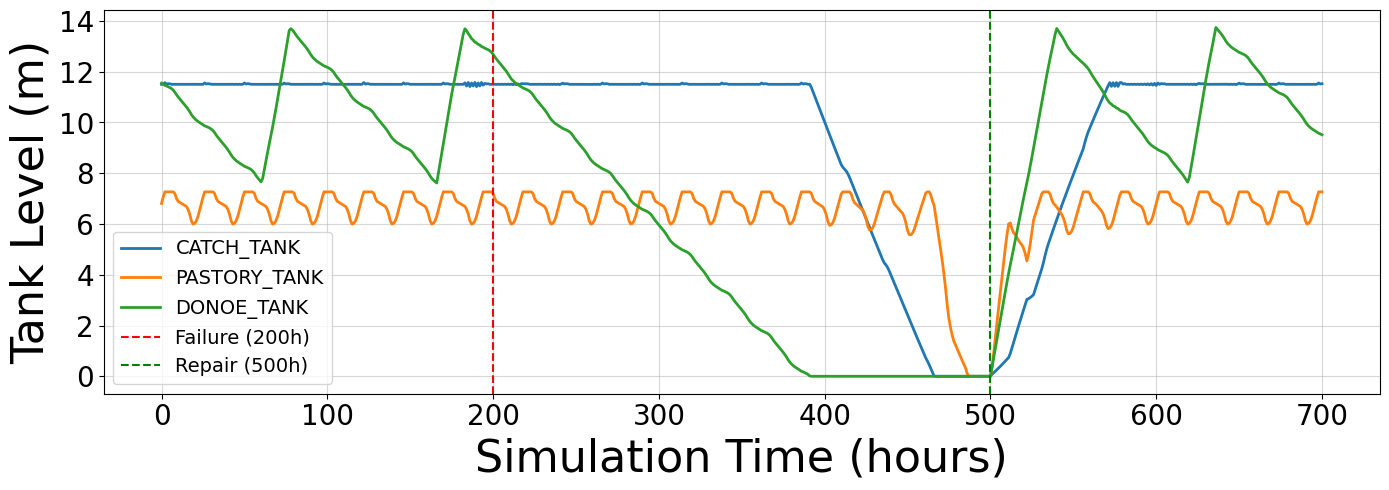}
        \caption{sttw\_50}
        \label{subfig:sttw50}
    \end{subfigure}
    \hfill 
    \begin{subfigure}[b]{0.48\textwidth}
        \centering        \includegraphics[width=\linewidth, height = 3.5cm]{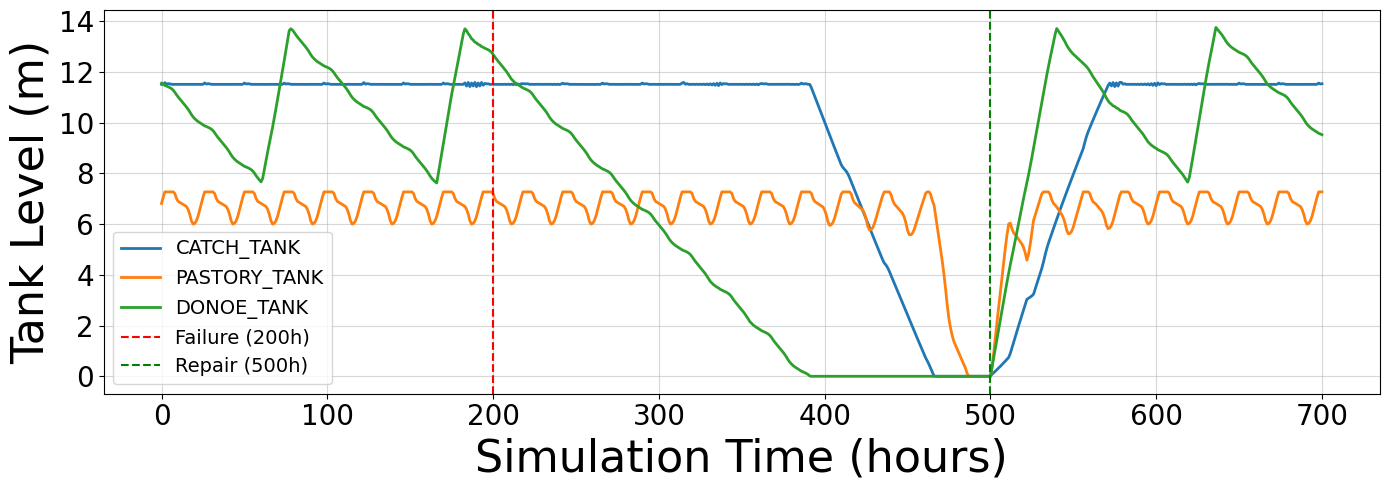}
        \caption{sttw\_87}
        \label{subfig:sttw87}
    \end{subfigure}
    
    \vspace{0.5cm}
    
    \begin{subfigure}[b]{0.48\textwidth}
        \centering        \includegraphics[width=\linewidth, height = 3.5cm]{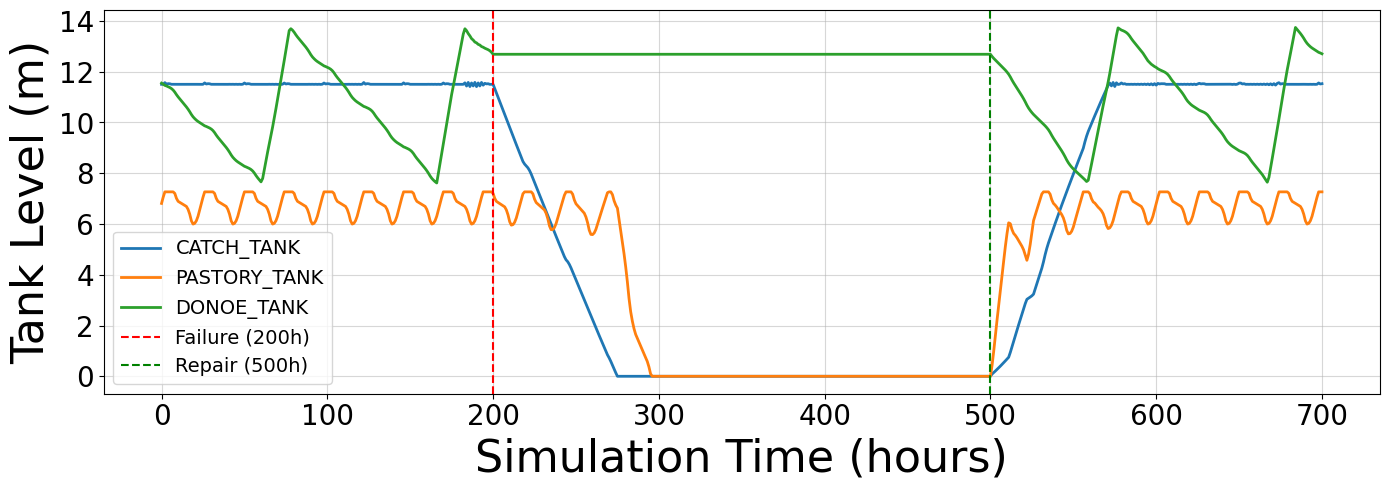}
        \caption{stte\_155}
        \label{subfig:stte155}
    \end{subfigure}
    \hfill 
    \begin{subfigure}[b]{0.48\textwidth}
        \centering        \includegraphics[width=\linewidth, height = 3.5cm]{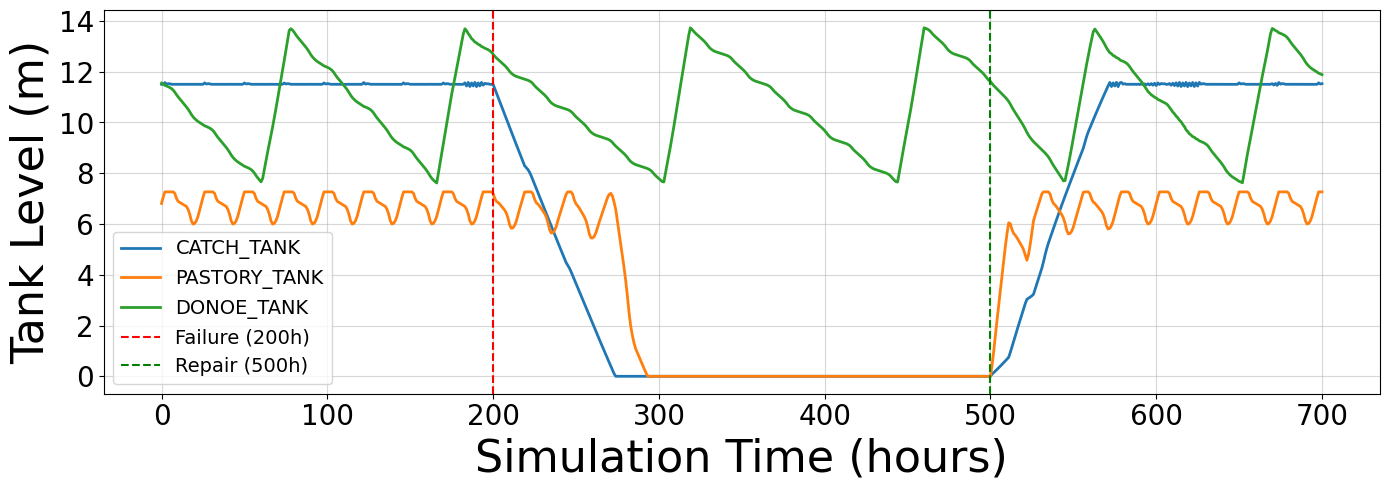}
        \caption{pipe23}
        \label{subfig:pipe23}
    \end{subfigure}   
    \caption{
        Tank level dynamics for four critical pipes: (a) sttw\_50, (b) sttw\_87, (c) stte\_155, and (d) pipe23. 
        Each simulation incorporates pipe failure at 200 hours and subsequent repair at 500 hours. 
        The trajectories illustrate topology-dependent responses and delayed hydraulic effects. 
        Tank depletion patterns vary depending on pipe location, and delayed responses indicate that 
        identical observable states may correspond to different underlying system conditions.
    }
    \label{fig:tank_dynamics_all_pipes}
\end{figure}

Figure~\ref{fig:tank_dynamics_all_pipes} illustrates the dynamics of tank levels under failure and repair scenarios for the four selected critical pipes. These trajectories reveal two key characteristics of the system. First, tank responses reflect the underlying network topology. For example, failure of $stte\_155$ inhibits outflow from the Donoe tank, resulting in sustained water levels, while failures of $sttw\_50$ and $sttw\_87$ lead to immediate depletion. In contrast, failure of $pipe23$ does not substantially affect the Donoe tank, since $pipe23$ primarily governs inter-island water transfer between STT and STJ.

Second, the system exhibits delayed responses following pipe failures. In particular, the Catch and Pastory tanks continue to supply water for a while before depletion begins. This behavior indicates that identical observable states may correspond to different underlying system conditions depending on the elapsed time since failure, which is consistent with the latent failure-time formulation.

The trajectories also motivate the inclusion of tank-level changes in the state representation. While the Catch tank remains relatively stable before depletion, the Donoe and Pastory tanks exhibit oscillatory behavior. Therefore, $\Delta L_t$ helps distinguish increasing, decreasing, and stable regimes that would not be captured by tank level alone.

Finally, we discretize each tank level into operational and non-operational states using thresholds derived from observed hydraulic dynamics and informed by operational expertise from the Virgin Islands Water and Power Authority. The operational thresholds are set to 6.1~m for the Catch tank, 5.2~m for the Pastory tank, and 7.6~m for the Donoe tank. We classify level changes using a threshold of 0.01~m, which filters numerical noise while preserving meaningful hydraulic variation.

\subsubsection{One-step Cost}

We analyze the WSA dynamics under failure and repair scenarios for the four critical pipes. For the purposes of this work, we define both the flow cost and the maintenance cost at \$10,000 per time step. Although precise cost figures are illustrative rather than empirically validated, this value reflects the order of magnitude of operational costs in island water systems—including emergency water delivery logistics and specialized marine labor required for submarine pipeline maintenance \citep{alderson_interdependent_2018}—and is set equal for both cost components to establish a neutral baseline for sensitivity analysis. The results are fictitious, since the actual cost data was not used. However, it provides a useful estimate of what an optimal repair policy would look like.

\begin{figure*}[htbp]
    \centering
    
    \begin{subfigure}[b]{0.48\textwidth}
        \centering
        \includegraphics[width=\linewidth, height = 3.5cm]{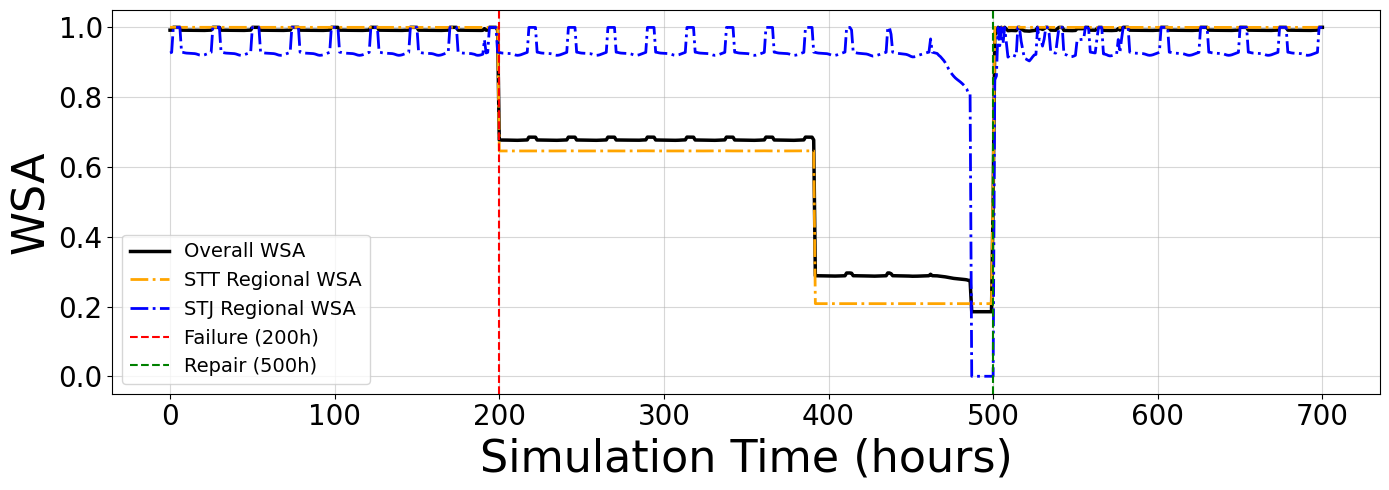}
        \caption{sttw\_50}
        \label{subfig:wsa_sttw50}
    \end{subfigure}%
    \hfill
    \begin{subfigure}[b]{0.48\textwidth}
        \centering
        \includegraphics[width=\linewidth, height = 3.5cm]{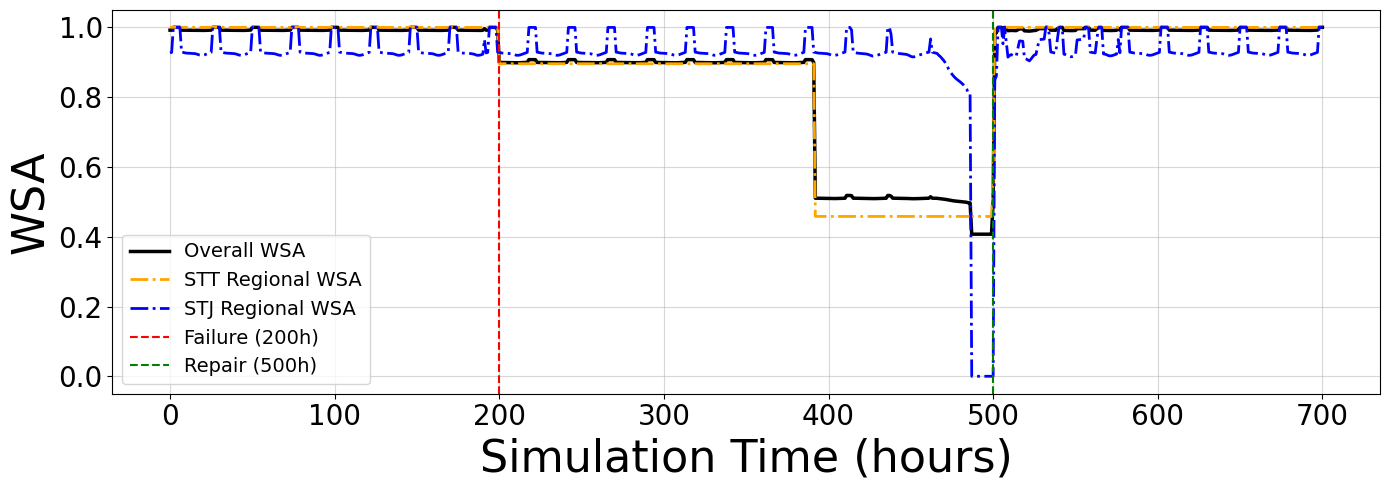}
        \caption{sttw\_87}
        \label{subfig:wsa_sttw87}
    \end{subfigure}
    
    \vspace{0.5cm} 
    
    \begin{subfigure}[b]{0.48\textwidth}
        \centering
        \includegraphics[width=\linewidth, height = 3.5cm]{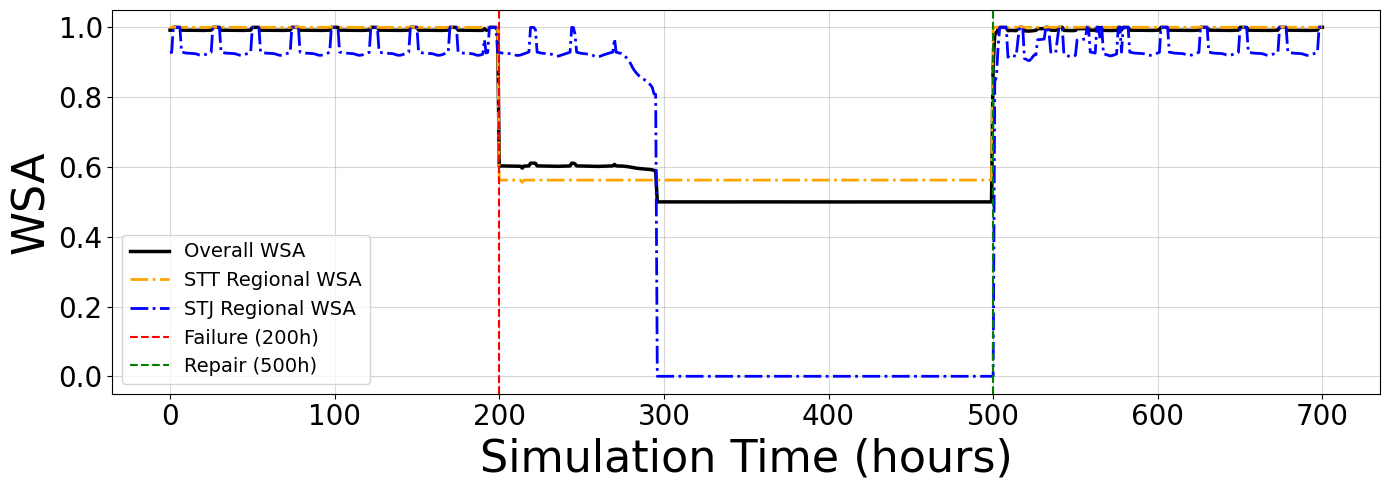}
        \caption{stte\_155}
        \label{subfig:wsa_stte155}
    \end{subfigure}%
    \hfill
    \begin{subfigure}[b]{0.48\textwidth}
        \centering
        \includegraphics[width=\linewidth, height = 3.5cm]{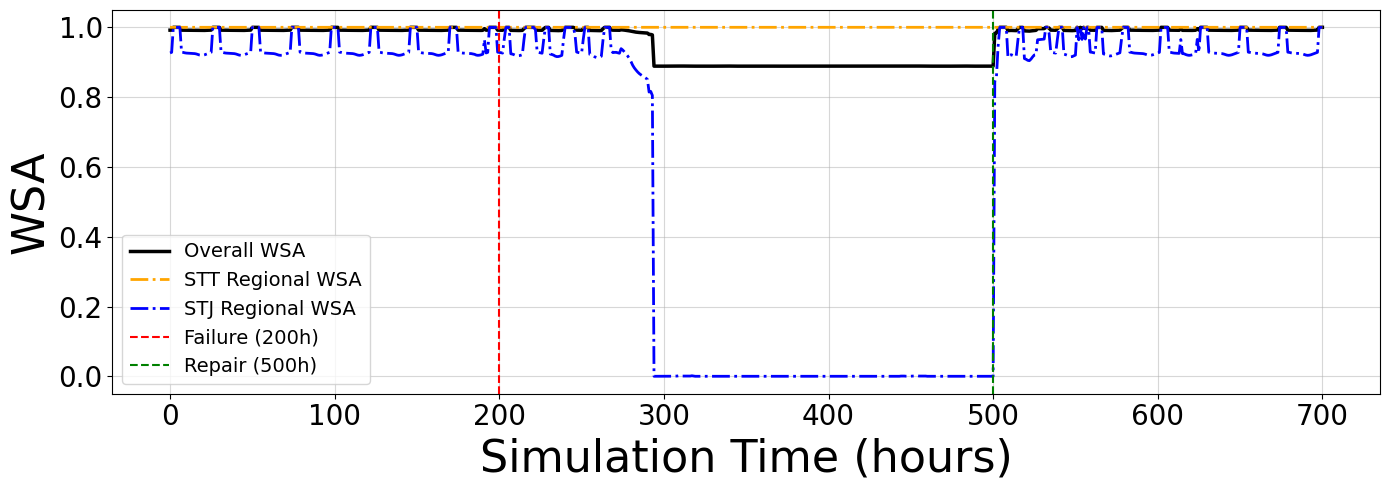}
        \caption{pipe23}
        \label{subfig:wsa_pipe23}
    \end{subfigure}
    
    \caption[WSA dynamics under failure and repair of four critical pipes]{
        WSA dynamics under failure at 200 hours and repair at 500 hours for four critical pipes: (a) sttw\_50, (b) sttw\_87, (c) stte\_155, and (d) pipe23. 
        Across all scenarios, WSA exhibits a delayed but abrupt degradation following failure due to storage depletion, and rapid recovery after repair.
    }
    \label{fig:wsa_dynamics_all_pipes}
\end{figure*}

Figure~\ref{fig:wsa_dynamics_all_pipes} presents the evolution of the overall and regional WSA over time. Across all pipe scenarios, we observe consistent structural patterns in service availability. Under normal operation, the WSA remains close to one, with minor oscillations at certain junctions due to local hydraulic dynamics. However, following pipe failure, service degradation does not occur immediately. Instead, WSA remains stable for a period due to storage buffering in tanks, after which a rapid and significant drop occurs once storage is depleted.

This delayed but abrupt degradation is a key characteristic of the system. The timing of the drop varies depending on the pipe and its topological role in the network, but the overall pattern remains consistent across all scenarios. In particular, failures in pipes that directly affect inter-regional connectivity (e.g., $pipe23$) or upstream supply (e.g., $sttw\_50$) lead to more severe and earlier degradation.

Following repair, WSA recovers rapidly to near-unity levels, even before full recovery of tank levels. This indicates that service availability is primarily governed by the ability to meet demand rather than by complete storage restoration.

These observations provide strong empirical support for the threshold-based flow cost formulation introduced in the Methods section. Since service degradation is negligible above a certain level but becomes critical once WSA drops below a threshold, a binary penalty structure effectively captures the operational impact of service loss. Based on the observed WSA dynamics, we set the service availability threshold to $\theta = 0.52$, which corresponds to the lower bound of oscillatory behavior under nominal conditions and distinguishes meaningful service disruptions from normal fluctuations.

In general, the consistency of these patterns across multiple pipe scenarios supports the generality of the proposed cost structure and validates its use in the subsequent MDP analysis.

\subsection{Discounted MDP Solution}

We define the initial daily failure probability, $p_{fail} = 0.05$, indicating a reasonable risk level. We later modify the value to assess policy robustness through sensitivity analysis. In addition, we assign the discount factor, $\gamma = 0.95$ to balance a long-term infrastructure perspective with the necessary convergence of the discounted MDP.

\subsubsection{Optimal Policy Structure}

Appendix~\ref{sec:appendix-b}, Table~\ref{tab:policy_integrated_full} presents the full integrated optimal policy table across the four critical pipe scenarios, while Table~\ref{tab:policy_representative} summarizes the representative policy patterns for readability and discussion. The results show that the optimal policy primarily selects repair actions when one or more tanks enter non-operational and decreasing states. In contrast, when tank levels remain operational and stable or increasing, the policy generally favors the \textit{DoNothing} action.

We observe that even when the system appears operational (e.g., OP|MAINT|OP|MAINT|OP|MAINT), the optimal policy selects the \textit{Repair} action. This indicates that the observable state does not fully capture the underlying system condition. In particular, latent factors such as the elapsed time since failure influence the future system dynamics. As a result, the network may appear normal while being in a degraded or pre-failure regime.

We further identify states that occur exclusively under specific pipe failure scenarios. For example, a state in which the Donoe tank remains operational while the Catch tank falls below its operational threshold is observed only when $pipe23$ fails. This exclusivity reflects the structural role of $pipe23$ as the sole inter-island connection, where its failure isolates downstream storage from upstream supply. This emergent property demonstrates that observable state transitions can identify localized failures, providing a direct basis for targeted inspection.

\begin{table*}[!ht]
\centering
\caption{Representative optimal policies selected from full optimal policy (Appendix~\ref{sec:appendix-b}, Table~\ref{tab:policy_integrated_full})}
\label{tab:policy_representative}
%\resizebox{\textwidth}{!}
{
\begin{tabular}{l l c c c c}
\toprule
\textbf{Scenario} & \textbf{State} & \textbf{$sttw\_50$} & \textbf{$sttw\_87$} & \textbf{$stte\_155$} & \textbf{$pipe23$} \\
\midrule
Normal & OP|MAINT|OP|MAINT|OP|MAINT & Repair & Repair  & Repair & DoNothing \\
Failure & OP|DEC|OP|DEC|OP|DEC & DoNothing & DoNothing & DoNothing & DoNothing \\
Failure & NOP|DEC|OP|DEC|OP|DEC & Repair & Repair & NA & NA \\
Failure & OP|DEC|NOP|DEC|OP|DEC & NA & NA & DoNothing & Repair  \\
Failure & NOP|DEC|OP|INC|OP|DEC & DoNothing & Repair & NA & NA \\
Repair & NOP|INC|NOP|DEC|NOP|DEC & DoNothing & DoNothing & NA & NA \\
Repair & OP|INC|OP|INC|OP|DEC & DoNothing & DoNothing & DoNothing & DoNothing \\
\bottomrule
\multicolumn{6}{l}{\footnotesize \textit{Note:} OP: operational, NOP: non-operational, INC: increase, DEC: decrease, MAINT: maintain.}
\end{tabular}%
}
\end{table*}

\subsubsection{Cost and Repair Patterns}
Table~\ref{tab:repair_benchmark} summarizes the number of valid states, repair ratio, and total expected cost across pipe scenarios. The valid states represent physically realizable state configurations generated by the hydraulic dynamics under each pipe scenario. The repair ratio is the proportion of states for which the optimal policy selects the \textit{Repair} action, reflecting how proactively the policy acts to restore the system under different failure conditions. Both the number of reachable states and the repair ratio differ substantially by pipe, indicating that the optimal policy is sensitive to the underlying network topology and the resulting system dynamics.

We observe that $sttw\_50$ and $sttw\_87$ are associated with a larger number of valid states and higher repair frequencies compared to $pipe23$. This suggests that failures in these pipes propagate through
a wider range of system configurations, leading to more frequent intervention decisions. In contrast, $stte\_155$ exhibits a smaller set of valid states and a lower repair frequency, yet incurs high costs when activated. This indicates that failures of $stte\_155$ occur in a more restricted region of state space but lead to disproportionately severe service degradation—a rare but high-impact failure mode.

\begin{figure}[!ht]
  \centering
  \includegraphics[width=0.7\linewidth, height=6cm]{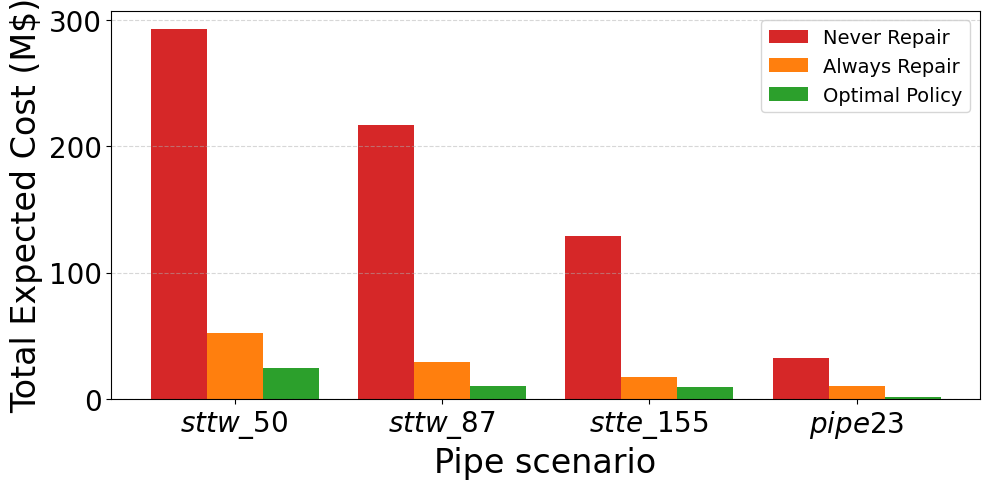}
  \caption{Total expected discounted cost (M\$) under three policies---Never Repair, Always Repair, and the optimal policy---across the four critical pipe scenarios. The optimal policy consistently achieves the lowest cost relative to Always and Never Repair policies.}
  \label{fig:benchmark_comparison}
\end{figure}

To quantify the value of state-dependent decision-making, we compare the optimal policy against two rule-based benchmarks: \textit{Always Repair} (select Repair in every reachable state) and \textit{Never Repair} (select DoNothing in every reachable state). Both benchmarks are evaluated using the same LP-based policy evaluation framework. Table~\ref{tab:repair_benchmark} and Figure~\ref{fig:benchmark_comparison} summarize and visualize the difference between the pipe scenarios.

Across all four pipe scenarios, the optimal policy substantially outperforms both benchmarks. The total expected cost under optimal policy is consistently the lowest, ranging from
\$1 M ($pipe23$) to \$24 M ($sttw\_50$). Relative to Always Repair, the optimal policy reduces the expected discounted cost by 44.35\% ($stte\_155$) to 86.67\% ($pipe23$), demonstrating that indiscriminate repair incurs substantial unnecessary expenditure. The gains are even more pronounced relative to Never Repair, where the cost reductions range from 91.79\% ($sttw\_50$) to 95.78\% ($pipe23$), reflecting the severe service degradation that accumulates without any maintenance intervention.

The heterogeneity across pipes is operationally significant. The large savings of the optimal policy relative to Always Repair for $pipe23$ (86.67\%) indicate that most observable states do not warrant repair under the baseline cost parameters—consistent with its low repair ratio of 16.7\%. In contrast, the
comparatively smaller savings for $stte\_155$ relative to Always Repair (44.35\%) reflect that failures of this pipe tend to occur in states where repair is more frequently optimal, given the severity of downstream service degradation. Together, these results demonstrate
that the LP-derived optimal policy captures non-trivial, state-dependent repair decisions that neither naïve benchmark can replicate, and that the economic benefit of adopting the optimal policy is substantial
across all pipe scenarios considered.

\begin{table}[!ht]
\centering
\caption{Repair frequency and expected discounted cost comparison}
\label{tab:repair_benchmark}
% \resizebox{\linewidth}{!}{
\begin{tabular}{l r r r r r}
\toprule
\multirow{2}{*}{\textbf{Pipe}}
  & \multicolumn{2}{c}{\textbf{Optimal Policy}}
  & \multicolumn{3}{c}{\textbf{Expected Cost (M\$)}} \\
\cmidrule(lr){2-3} \cmidrule(lr){4-6}
  & \textbf{Valid states}
  & \textbf{Repair ratio}
  & \textbf{Optimal}
  & \textbf{AR}
  & \textbf{NR} \\
\midrule
$sttw\_50$  & 62 & 25.8\% & 24 & 52 & 293 \\
$sttw\_87$  & 62 & 29.0\% & 10 & 29 & 217 \\
$stte\_155$ & 40 & 22.5\% &  9 & 17 & 129 \\
$pipe23$    & 48 & 16.7\% &  1 & 10 &  32 \\
\bottomrule
\multicolumn{6}{l}{\footnotesize
  \textit{Note:} AR\,=\,Always Repair; NR\,=\,Never Repair.}
\end{tabular}
% }
\end{table}

% These results highlight that each pipe induces a different structure of feasible system transitions, reflecting its topological role in the network. Consequently, repair frequency alone does not fully characterize pipe criticality. Instead, both the likelihood of occurrence (frequency) and the severity of impact (cost) must be jointly considered to evaluate infrastructure importance.

\FloatBarrier

\subsection{Sensitivity Analysis}

We conduct a sensitivity analysis on the representative pipe scenario ($pipe23$). Since the primary objective of this analysis is to characterize structural policy transitions rather than compare operational importance across all pipes, we focus on a single representative critical pipe to avoid redundant analyses. The analysis of the discount factor reveals that the variation of $\gamma$ mainly changed the magnitude of the total expected cost, while the structure of the optimal policy remained unchanged. This suggests that the qualitative structure of the optimal policy is robust to the choice of discount factor.

\subsubsection{Expected Cost Sensitivity}

Figure~\ref{subfig:sensi_cost_pipe23} illustrates the total expected cost as a function of $\lambda$ and $p_{fail}$. As $p_{fail}$ increases, the sensitivity of the expected cost to $\lambda$ becomes markedly more pronounced. When $p_{fail}$ is extremely low (e.g., $p_{fail}=0.0001$), the total expected cost remains largely invariant across the range of $\lambda$. This indicates that maintenance decisions exert a marginal influence when disruptions are extremely rare.

In contrast, higher values of $p_{fail}$ precipitate distinct regime changes in the expected cost structure. Appendix C, Figures~\ref{fig:cost decomposition_vs_lambda_p_failed=0.01_pipe23}, \ref{fig:cost decomposition_vs_lambda_p_failed=0.05_pipe23}, \ref{fig:cost decomposition_vs_lambda_p_failed=0.1_pipe23} show that increasing $\lambda$ forces the maintenance cost component to outweigh the incentive for immediate repair, causing the optimal policy to defer or eliminate Repair actions. Once the policy transitions toward low- or no-repair regimes, the flow cost escalates rapidly due to the accumulation of service degradation and prolonged operation under failed conditions.

These results indicate that disruption frequency and the relative maintenance cost jointly dictate the effectiveness of maintenance policies. Heightened disruption risks prompt an earlier transition to low- or no-repair regimes, thereby driving flow penalties to dominate the overall expected costs.
Therefore, there is a need to tailor maintenance strategies to operational risk environments such as infrastructure aging, natural disasters, or adversarial disruptions, under which the probability of system failure may vary substantially.

%\begin{figure} 
%    \centering
%    \includegraphics[width=\linewidth]{sensi_totalcost_vs_lambda_across_different_p_failed_pipe23.png}
%    \caption[Total expected cost vs $\lambda$ across the probability of pipe23 failure ($p_{fail}$)]{Total expected cost vs $\lambda$ across the probability of pipe23 failure ($p_{fail}$).}
%\label{fig:totalcost_vs_lambda_across_different_p_failed_pipe23}
%\end{figure}

\subsubsection{Policy Structure Sensitivity (Repair Ratio)}

\begin{figure}[!ht]
    \centering
    \begin{subfigure}[b]{\linewidth}
        \centering
        \includegraphics[width=0.6\linewidth]{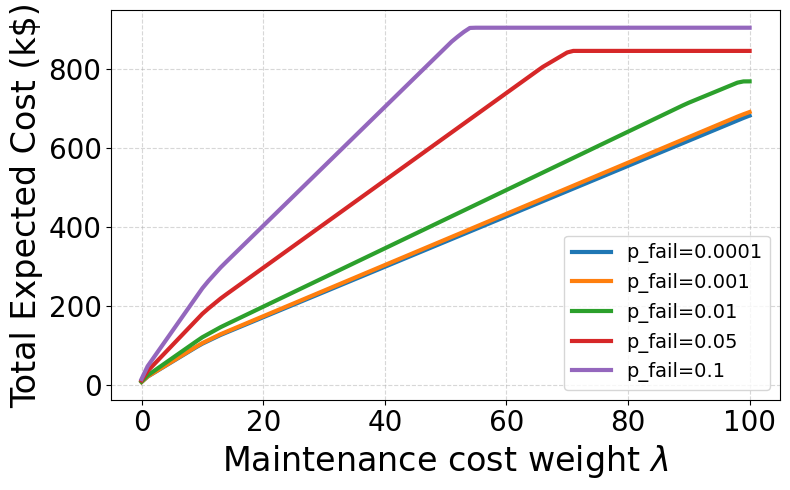}
        \caption{Total expected cost}
        \label{subfig:sensi_cost_pipe23}
    \end{subfigure}
    
    \vspace{0.5cm} 
    
    \begin{subfigure}[b]{\linewidth}
        \centering
        \includegraphics[width=0.63\linewidth]{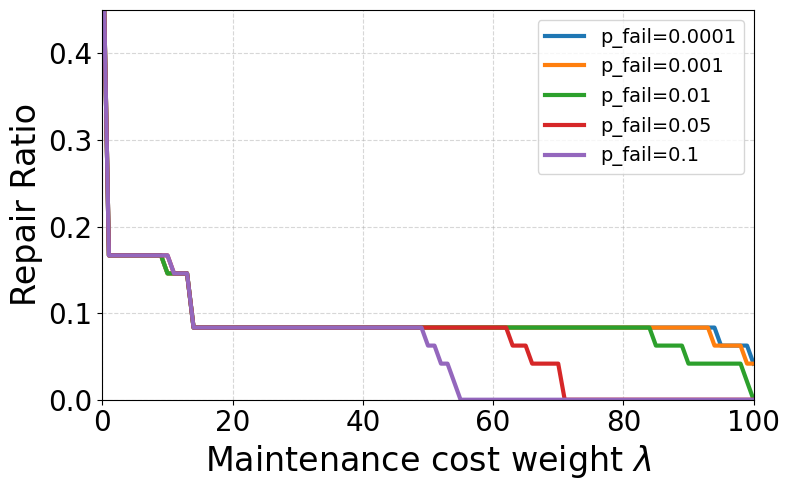}
        \caption{Repair ratio}
        \label{subfig:sensi_repair_pipe23}
    \end{subfigure}
    
    \caption[Sensitivity analysis for pipe23]{Sensitivity analysis across the probability of pipe23 failure ($p_{fail}$) and the relative maintenance cost weight ($\lambda$). (a) Total expected cost. (b) Repair ratio. As $p_{fail}$ increases, the optimal policy becomes more sensitive to $\lambda$, accelerating the transition toward low-intervention regimes.}
    \label{fig:sensitivity_analysis_pipe23}
\end{figure}

Figure~\ref{subfig:sensi_repair_pipe23} suggests how the repair ratio responds to varying $\lambda$ and $p_{fail}$. When $p_{fail}$ increases, the threshold value of $\lambda$ at which Repair actions substantially decrease or disappear changes to lower values. This indicates that, under more frequent disruptions, the optimal policy exhibits heightened sensitivity to maintenance cost, thereby accelerating the transition toward conservative, low-intervention regimes.

Conversely, when $p_{fail}$ is exceptionally low ($p_{fail}=0.0001$), the repair ratio gradually decays and avoids converging to zero, even at high values of $\lambda$. In such scenarios, repair primarily functions as a preventive action that mitigates rare but high-impact trajectories, rather than responding to frequent disruptions. Despite the rarity of failures, the long-run discounted costs of persistent degraded states remain substantial enough to justify selective repairs.

In general, these results indicate that while $\lambda$ dictates the existence and placement of policy regime boundaries, $p_{fail}$ determines how early they emerge. Higher failure probabilities drive faster transitions to no-repair regimes, whereas lower probabilities sustain repair actions across a broader spectrum of maintenance costs. Ultimately, these findings underscore the necessity of jointly evaluating disruption risks and economic trade-offs to design resilient infrastructure maintenance policies.

\FloatBarrier

\section{Discussion}\label{sec:discussion}

Our results have broad implications for optimal maintenance decisions for WDNs and critical infrastructure networks. Key implications of this work include:

\begin{itemize}
    \item \textbf{MDPs and hydraulic simulation:} our results demonstrate the need to verify and design MDP-based decision models with a hydraulic simulation of system dynamics. Otherwise, the state-action space may not retain the Markov property and resulting decisions may be suboptimal.
    \item \textbf{Non-intuitive optimal policies due to network effects:} WDN network operations, including tank storage, pipe settings, and demand patterns, combine such that pipes that may appear more important than others (i.e., upstream) are actually less impactful on system operations.
    \item \textbf{"State Fingerprinting" via indirect sensing:} Some operational states (i.e., tank levels and gradients) are only possible given specific pipe failure configurations. This creates an indirect and novel way to anticipate potential water outages and optimal policies in data-poor environments.
\end{itemize}

\noindent We provide further discussion of these and other implications of this work.

\subsection{MDP Maintenance Modeling of WDNs}
Unlike many previous studies that adopt the Markov assumption without empirical verification, our analysis demonstrates that only specific temporal discretizations approximately satisfy the Markov property for the WDN dynamics considered in this study. This is an important contribution because maintenance policies derived under inappropriate time discretizations may inadequately capture the temporal evolution of hydraulic states and service degradation. This result is meaningful not just for our integrated MDP model, but other reinforcement learning models and studies that assume Markovian dynamics in WDNs. 

Furthermore, the proposed framework can be generalized and automated for future or newly designed water systems that exist only in simulation or model form. In general, hydraulic simulation models like WNTR are often used in system design and planning stages of WDNs, rather than on-line operations. Since the framework integrates hydraulic simulation with MDP-based decision-making, multiple candidate network designs can be evaluated according to their expected repair policies, operational resilience, implementation costs, and maintenance difficulty prior to their implementation in a real system.

Another practical contribution of this work is that the framework relies only on operationally observable variables, such as tank levels, rather than requiring direct sensing of underground pipe conditions. This is particularly important in data-sparse environments such as the USVI, where utilities may lack extensive sensing infrastructure or SCADA coverage. The results suggest that physically observable hydraulic states can still provide sufficient information to support maintenance planning, failure diagnosis, and long-term infrastructure management.

\subsection{Topological Vulnerability Beyond Geographic Intuition}

The results reveal that geographic proximity to the water source does not solely determine the impacts of a pipe failure. In particular, in every state that is reachable under both the $stte\_155$ and $sttw\_87$ failure scenarios, the expected cost per-state under $stte\_155$ consistently exceeds that under $sttw\_87$ (Table~\ref{tab:policy_integrated_full}) despite $sttw\_87$ being located further upstream and closer to the reservoir. This counterintuitive outcome arises from the network topology and storage dynamics. When $sttw\_87$ fails, the Donoe tank acts as a buffer, temporarily sustaining downstream supply. In contrast, failure of $stte\_155$ directly disrupts water delivery to demand-critical regions, resulting in more severe service degradation.

Our MDP framework successfully captures these hidden topological vulnerabilities, which simple geographic reasoning often obscures. Consequently, infrastructure managers must incorporate system dynamics and storage buffer effects when prioritizing inspection and maintenance interventions. Practically, this finding motivates managers to reorder inspection priorities, shifting away from conventional distance-based or age-based heuristics. We recommend segmenting the network by \textit{functional consequence}. Managers should schedule more frequent condition assessments for pipes that bypass existing storage buffers (such as $stte\_155$). In contrast, pipes whose failures are partially absorbed by upstream tanks (such as $sttw\_87$) can tolerate longer inspection intervals without imposing a disproportionate service risk.

\subsection{State Fingerprinting as a Virtual Sensing Mechanism}

We observe that certain tank state configurations arise exclusively under specific pipe failure scenarios. This implies that the system exhibits a form of \emph{state fingerprinting}, where observable tank dynamics encode information about hidden infrastructure conditions.

In the context of the USVI, where pipe-level sensing is largely unavailable and infrastructure monitoring resources are limited, this property is particularly valuable. In such data-sparse environments, these distinctive state patterns effectively serve as a virtual sensing mechanism. By observing the system state, decision-makers can infer the likely location of failures without requiring direct inspection, enabling more informed inspection planning and prioritization.

This capability is especially important for long-term infrastructure management. It allows operators to track system conditions indirectly, anticipate potential degradation, and integrate inspection and repair decisions within a unified planning framework. As a result, the approach supports both operational responsiveness and strategic cost minimization in resource-constrained settings.

Concretely, this state fingerprinting property distills the high-dimensional MDP policy into intuitive, visual-based diagnostic protocols for operators. For example, the policy prescribes immediate \textit{Repair} for $pipe23$ exclusively when the Catch tank drops below its threshold while Donoe remains operational. Although suspecting a fault in the sole submarine pipeline under these conditions may seem intuitively obvious, the model's autonomous codification of this physical intuition validates the framework's reliability. Furthermore, this diagnostic capability extends to more ambiguous subnetworks; failures in the western STT corridor ($sttw\_50$ or $sttw\_87$) are uniquely isolated to states where Donoe is non-operational and decreasing, while Catch and Pastory remain operational. Ultimately, these fingerprint-based dispatch rules translate complex optimal policies into actionable emergency response protocols, providing a mathematically backed decision aid that relies solely on simple tank readings without requiring on-line computation.

\section{Conclusion}\label{sec:conclusion}

This study develops a repair-oriented inspection and maintenance decision framework for WDN in data-sparse environments, motivated by the operational challenges of the USVI system. In the absence of pipe sensors and complete system information, the framework aims not only to minimize life-cycle maintenance costs while maintaining service availability but also to provide actionable insights for inspection and operational decision-making.

We formulate the problem as an MDP and solve it using a linear programming approach. To ensure modeling validity, we empirically assess the Markov property and identify an appropriate temporal discretization. The proposed approach integrates repair-oriented decision-making with high-fidelity hydraulic simulation based on physics using WNTR, enabling policy optimization grounded in realistic system dynamics.

The results reveal structured and interpretable optimal policies that depend on system state and pipe scenario. In particular, we identify heterogeneous failure characteristics across pipes, including frequent-impact and rare-but-high-impact behaviors. Furthermore, we show that certain system states uniquely correspond to specific pipe failures, enabling a form of state-based failure identification. This state fingerprint effect provides a virtual sensing capability that is especially valuable in environments with limited monitoring infrastructure.

Importantly, the proposed framework captures delayed system responses, sequential effects, and disruption-driven dynamics that are difficult to represent using traditional maintenance models. By incorporating latent temporal dependencies within the transition structure, the model preserves physical realism while maintaining computational tractability and policy interpretability.

Ultimately, this work demonstrates how decision-makers can leverage system-level dynamics to support inspection planning and maintenance decisions under uncertainty, providing a practical and scalable approach for infrastructure resilience management in resource-constrained settings.

While this study provides a robust foundation for maintenance decision-making, several limitations warrant acknowledgment and suggest directions for future research. First, the current framework evaluates critical pipes independently rather than modeling simultaneous multi-component failures. Although expanding the state space to accommodate concurrent failures would significantly increase computational complexity---potentially introducing the curse of dimensionality---it could reveal compounding vulnerabilities and synergistic policy structures. Second, transition probabilities are currently estimated using a counting-based approach that assumes uniform failure distributions within time steps. Future work could enhance this by integrating Bayesian estimation methods or more sophisticated age-dependent stochastic degradation models, such as Weibull or geometric distributions. Finally, the action space could be broadened beyond the binary choice of \textit{DoNothing} and \textit{Repair}. Introducing varying degrees of intervention—such as distinguishing between imperfect maintenance (e.g., patching) and perfect repair (e.g., full replacement)—would enrich the analysis of trade-offs between immediate capital expenditure and long-term system resilience under more realistic operational constraints.

% \clearpage

\bibliographystyle{abbrvnat}
\bibliography{refs}  %%% Uncomment this line and comment out the ``thebibliography'' section below to use the external .bib file (using bibtex) .

%%% Uncomment this section and comment out the \bibliography{references} line above to use inline references.
% \begin{thebibliography}{1}

% 	\bibitem{kour2014real}
% 	George Kour and Raid Saabne.
% 	\newblock Real-time segmentation of on-line handwritten arabic script.
% 	\newblock In {\em Frontiers in Handwriting Recognition (ICFHR), 2014 14th
% 			International Conference on}, pages 417--422. IEEE, 2014.

% 	\bibitem{kour2014fast}
% 	George Kour and Raid Saabne.
% 	\newblock Fast classification of handwritten on-line arabic characters.
% 	\newblock In {\em Soft Computing and Pattern Recognition (SoCPaR), 2014 6th
% 			International Conference of}, pages 312--318. IEEE, 2014.

% 	\bibitem{hadash2018estimate}
% 	Guy Hadash, Einat Kermany, Boaz Carmeli, Ofer Lavi, George Kour, and Alon
% 	Jacovi.
% 	\newblock Estimate and replace: A novel approach to integrating deep neural
% 	networks with existing applications.
% 	\newblock {\em arXiv preprint arXiv:1804.09028}, 2018.

% \end{thebibliography}

\clearpage

\appendix
\section*{Appendix}
\section{Markov Property Validation}\label{sec:appendix-a}

\subsection{Summary of Optimal Time-Step Selection}

Table~\ref{tab:markov_summary} indicates that multiple time steps satisfy the Markov property based on the paired t-test (p-value > 0.05) across all pipe scenarios. Among these, 46 hours consistently appear as a robust candidate across all pipes, tanks, and operational scenarios.

In some failure scenarios, particularly when tank levels become fully depleted, transition comparisons cannot be computed due to the absence of variability in the state trajectories. We exclude these cases from the statistical evaluation, as they do not provide meaningful information about Markovian dependence.

Focusing on scenarios with informative system dynamics, the 46-hour time step exhibits consistently high p-values and low improvement ratios across all configurations. Therefore, we select 46-hour as a unified discretization that balances statistical validity and modeling consistency across heterogeneous pipe failure conditions.

\begin{table}[!ht]
\centering
\caption{Summary of Markov test results across pipes, scenarios, and tanks}
\label{tab:markov_summary}
\begin{tabular}{l r r}
\toprule
\textbf{Pipe} & \textbf{Consistent Passing Time Steps} & \textbf{Selected Time Step} \\
\midrule
$sttw\_50$  & 29h, 33h, 36h, 44h, 45h, 46h & \textbf{46h} \\
$sttw\_87$  & 26h, 29h, 42h, 43h, 44h, 46h & \textbf{46h} \\
$stte\_155$ & 38h (strong), 46h (robust)   & \textbf{46h} \\
$pipe23$    & 10h, 26h, 33h, 46h           & \textbf{46h} \\
\bottomrule
\end{tabular}
\end{table}

\subsection{Representative Markov Test Results}

\subsubsection{Markov-order Comparison}

\begin{figure}[!ht]
\centering
\includegraphics[width=0.7\linewidth]{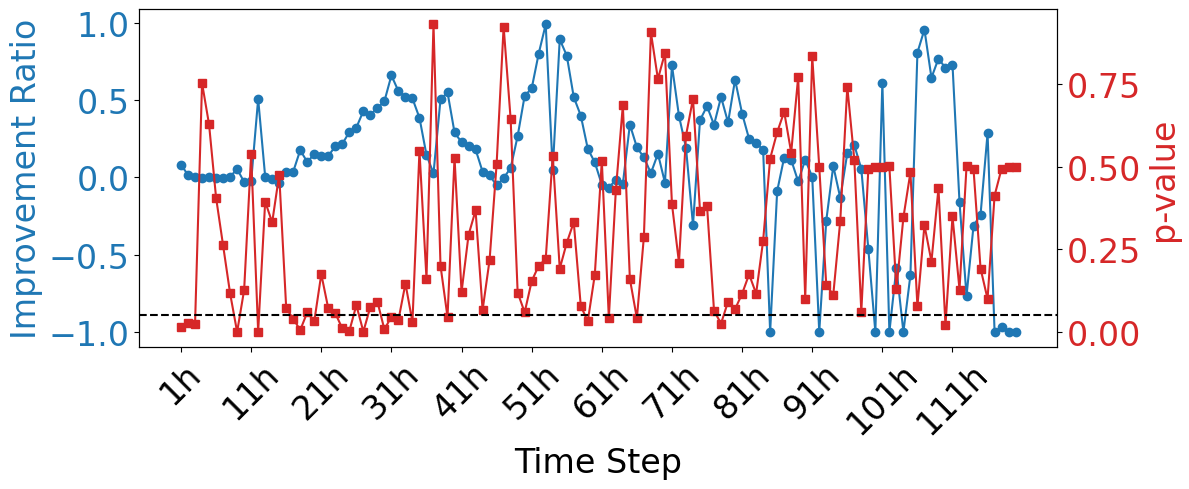}
\caption{Comparison of first- and second- order Markov models under the functional scenario for the Donoe tank.}
\label{fig:markov_order_2}
\end{figure}

\begin{figure}[!ht]
\centering
\includegraphics[width=0.7\linewidth]{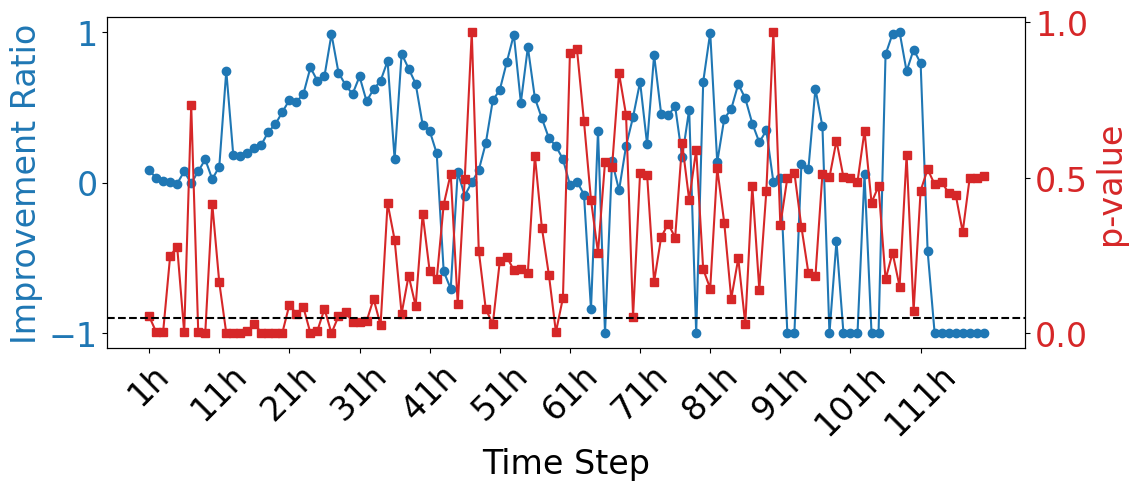}
\caption{Comparison of first- and third- order Markov models under the functional scenario for the Donoe tank.}
\label{fig:markov_order_3}
\end{figure}

Figures~\ref{fig:markov_order_2} and \ref{fig:markov_order_3} show that incorporating higher-order information does not significantly improve predictive performance across most time steps.

\FloatBarrier

\subsubsection{Tank Comparison}

\begin{figure}[!ht]
\centering
\includegraphics[width=0.7\linewidth]{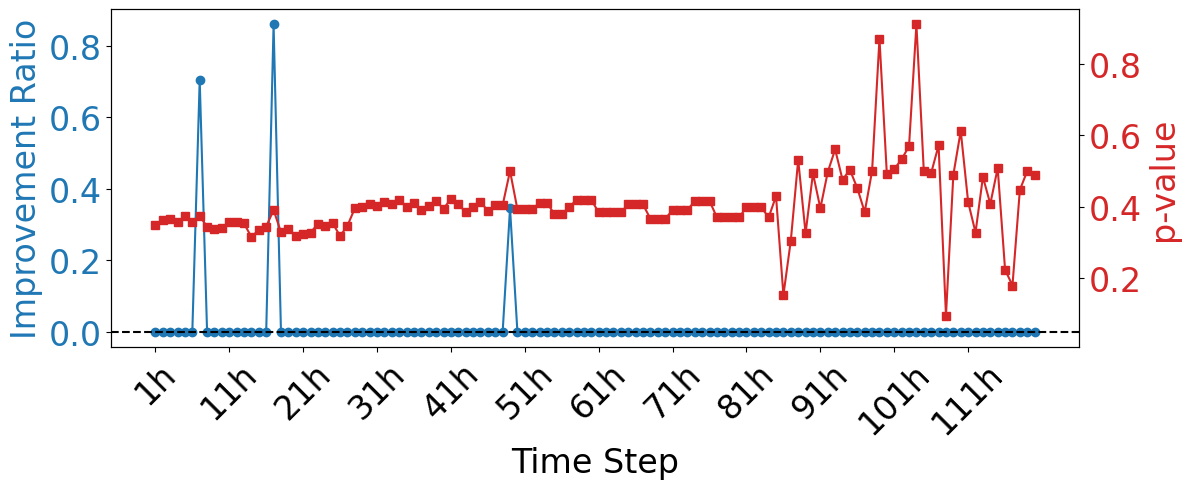}
\caption{Comparison of Markov model performance across tanks under the $sttw\_50$ failure scenario (Donoe).}
\label{fig:tank_comparison_donoe}
\end{figure}

\begin{figure}[!ht]
\centering
\includegraphics[width=0.7\linewidth]{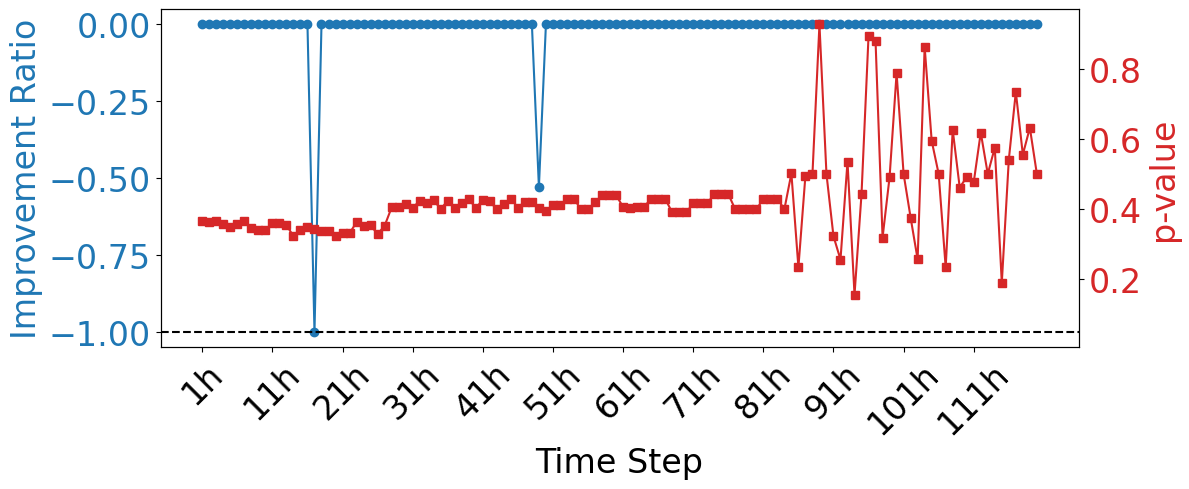}
\caption{Comparison of Markov model performance across tanks under the $sttw\_50$ failure scenario (Catch).}
\label{fig:tank_comparison_catch}
\end{figure}

Figures~\ref{fig:tank_comparison_donoe} and \ref{fig:tank_comparison_catch} indicate that the Donoe tank exhibits more variability and sensitivity to higher-order effects compared to the Catch tank.

\FloatBarrier

\subsubsection{Scenario Comparison}

\begin{figure}[!ht]
\centering
\includegraphics[width=0.7\linewidth]{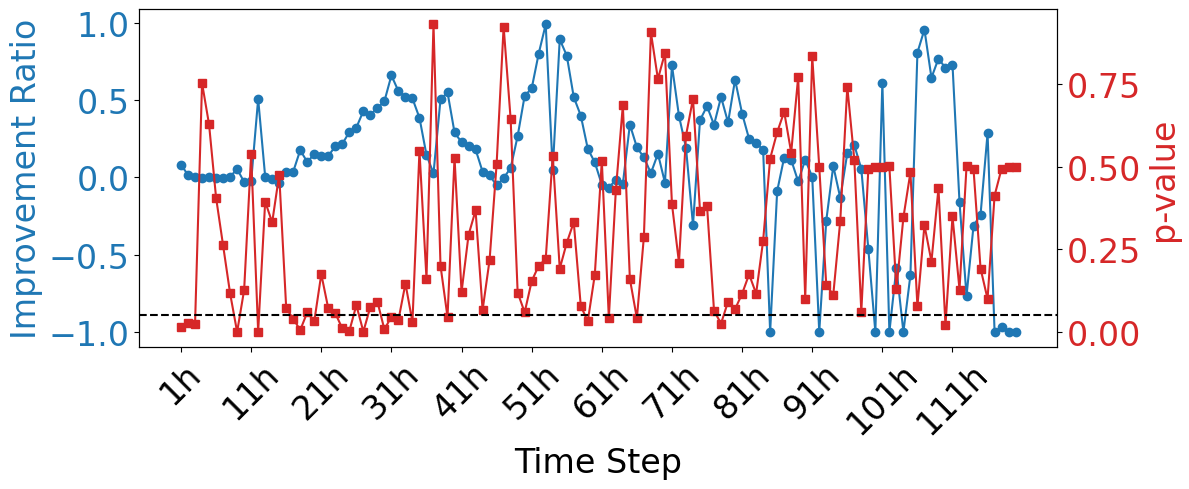}
\caption{Comparison of Markov model performance across scenarios (functional) for the Donoe tank.}
\label{fig:scenario_comparison_func}
\end{figure}

\begin{figure}[!ht]
\centering
\includegraphics[width=0.7\linewidth]{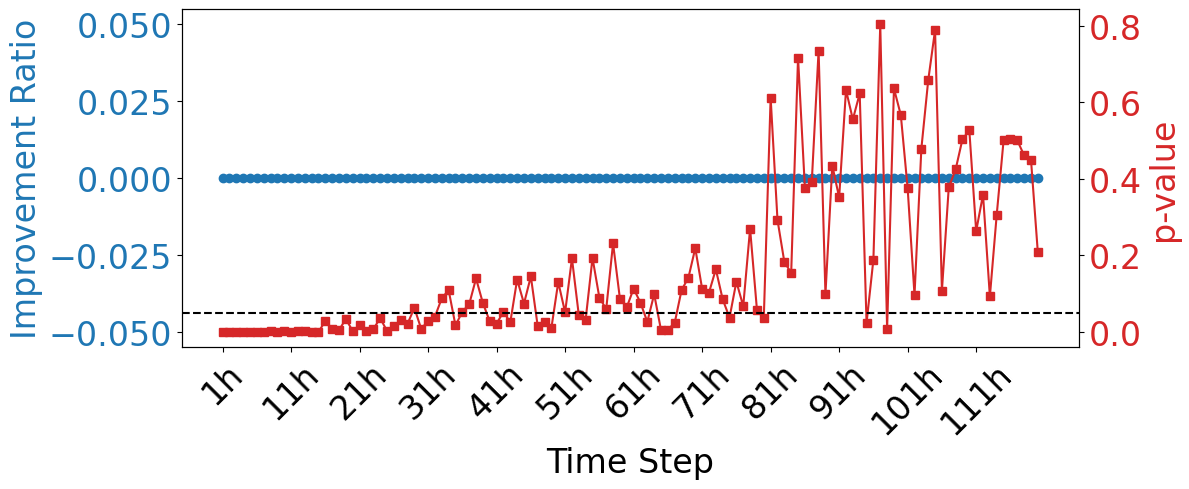}
\caption{Comparison of Markov model performance across scenarios (failure) for the Donoe tank.}
\label{fig:scenario_comparison_fail}
\end{figure}

\begin{figure}[!ht]
\centering
\includegraphics[width=0.7\linewidth]{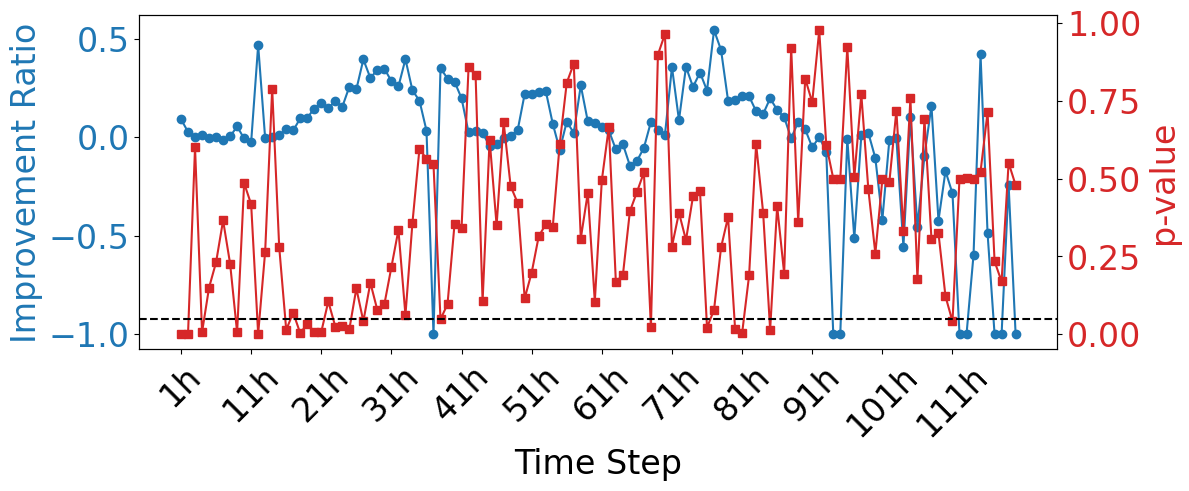}
\caption{Comparison of Markov model performance across scenarios (repair) for the Donoe tank.}
\label{fig:scenario_comparison_repair}
\end{figure}

Figures~\ref{fig:scenario_comparison_func}, \ref{fig:scenario_comparison_fail}, and \ref{fig:scenario_comparison_repair} show that the predictive structure varies significantly across system regimes, highlighting the importance of scenario-specific dynamics.

\FloatBarrier

\section{Additional Detailed Sensitivity Analysis on Cost Decomposition}

\begin{figure}[!ht]
    \centering
    \includegraphics[width=0.65\linewidth]{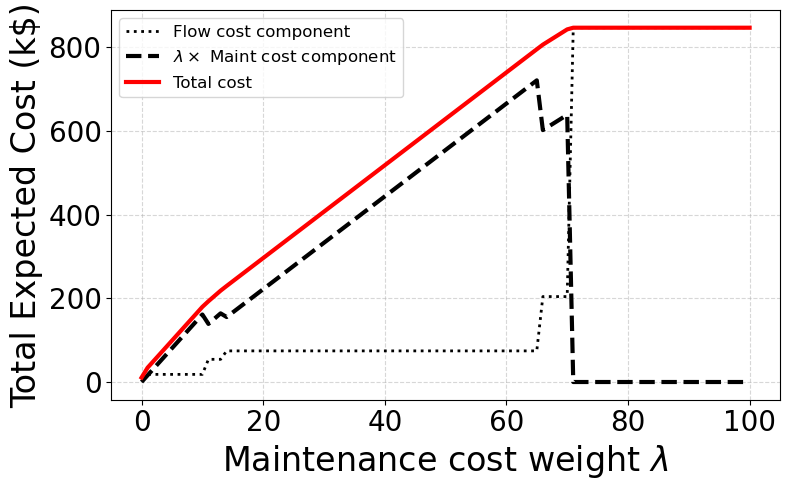}
    \caption[Cost decomposition vs $\lambda$ if the probability of pipe23 failure is $p_{fail}=0.05$]{Cost decomposition vs $\lambda$ if the probability of pipe23 failure is $p_{fail}=0.05$.}
    \label{fig:cost decomposition_vs_lambda_p_failed=0.05_pipe23}
\end{figure}

\begin{figure}[!ht]
    \centering
    \includegraphics[width=0.65\linewidth]{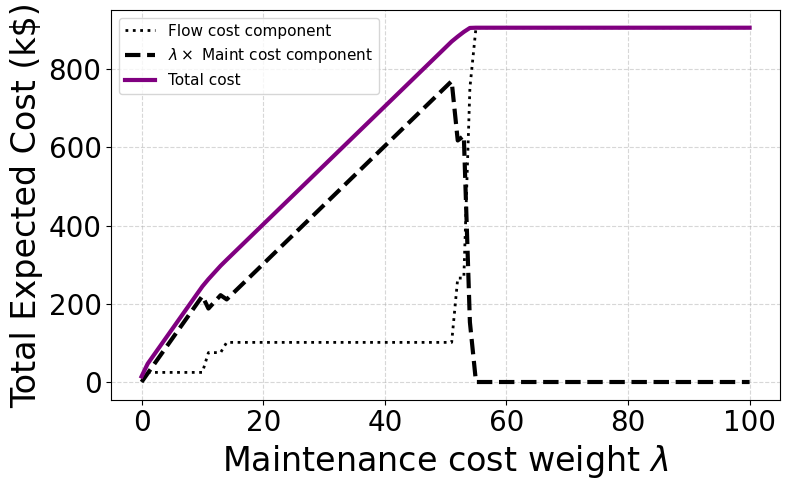}
    \caption[Cost decomposition vs $\lambda$ if the probability of pipe23 failure is $p_{fail}=0.1$]{Cost decomposition vs $\lambda$ if the probability of pipe23 failure is $p_{fail}=0.1$.}
    \label{fig:cost decomposition_vs_lambda_p_failed=0.1_pipe23}
\end{figure}

\begin{figure}[!ht]   
    \centering
    \includegraphics[width=0.65\linewidth]{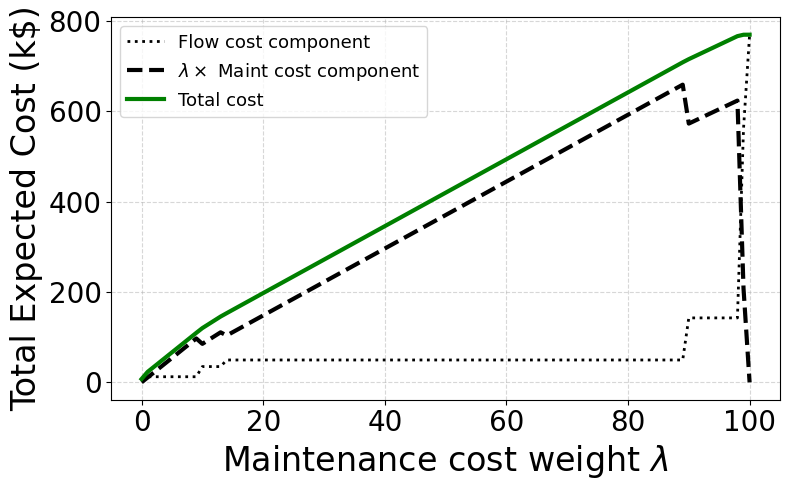}
    \caption[Cost decomposition vs $\lambda$ if the probability of pipe23 failure is $p_{fail}=0.01$]{Cost decomposition vs $\lambda$ if the probability of pipe23 failure is $p_{fail}=0.01$.}
    \label{fig:cost decomposition_vs_lambda_p_failed=0.01_pipe23}
\end{figure}

Figures~\ref{fig:cost decomposition_vs_lambda_p_failed=0.05_pipe23}, \ref{fig:cost decomposition_vs_lambda_p_failed=0.1_pipe23}, and 
\ref{fig:cost decomposition_vs_lambda_p_failed=0.01_pipe23} present the decomposition of the total expected cost into flow cost and maintenance cost across varying values of $\lambda$ from 0 to 100. As $p_{fail}$ increases, the transition from maintenance-cost-dominated regimes to flow-cost-dominated regimes occurs at smaller values of $\lambda$. This behavior reflects earlier suppression of Repair actions under higher disruption risk, which causes prolonged degraded operation and rapidly increasing service-loss penalties.

\clearpage
\onecolumn

\section{Integrated Policy Table}\label{sec:appendix-b}

\noindent
Table~\ref{tab:policy_integrated_full} reports the integrated optimal policy across the four critical pipe scenarios under the 46-hour MDP discretization. The state order is Donoe level, Donoe delta, Catch level, Catch delta, Pastory level, and Pastory delta. NA indicates that the state is not observed or not applicable for the corresponding pipe scenario.

\scriptsize
\setlength{\tabcolsep}{3pt}
\renewcommand{\arraystretch}{0.92}

\begin{longtable}{>{\raggedright\arraybackslash}p{0.29\textwidth} >{\centering\arraybackslash}p{0.16\textwidth} >{\centering\arraybackslash}p{0.16\textwidth} >{\centering\arraybackslash}p{0.16\textwidth} >{\centering\arraybackslash}p{0.16\textwidth}}
\caption{Integrated optimal policy for all four critical pipes (46-hour time step)}
\label{tab:policy_integrated_full}\\

\toprule
\textbf{State (Donoe level | Donoe delta | Catch level | Catch delta | Pastory level | Pastory delta)} & \textbf{$sttw\_50$} & \textbf{$sttw\_87$} & \textbf{$stte\_155$} & \textbf{$pipe23$} \\
\midrule
\endfirsthead

\toprule
\textbf{State} & \textbf{$sttw\_50$} & \textbf{$sttw\_87$} & \textbf{$stte\_155$} & \textbf{$pipe23$} \\
\midrule
\endhead

\midrule
\multicolumn{5}{r}{\textit{Continued on next page}}\\
\midrule
\endfoot

\bottomrule
\endlastfoot

\midrule
\multicolumn{5}{c}{\textbf{Functional-regime States}}\\
\midrule
OP|DEC|OP|MAINT|OP|DEC & DoNothing (480,263) & DoNothing (170,001) & DoNothing (210,276) & DoNothing (24,267) \\
OP|DEC|OP|MAINT|OP|INC & DoNothing (444,038) & DoNothing (168,049) & DoNothing (210,276) & DoNothing (23,830) \\
OP|DEC|OP|MAINT|OP|MAINT & DoNothing (472,564) & DoNothing (175,553) & DoNothing (210,276) & DoNothing (24,877) \\
OP|INC|OP|MAINT|OP|DEC & DoNothing (442,399) & DoNothing (161,521) & DoNothing (210,276) & DoNothing (23,475) \\
OP|INC|OP|MAINT|OP|INC & DoNothing (426,374) & DoNothing (161,632) & DoNothing (210,276) & DoNothing (23,558) \\
OP|INC|OP|MAINT|OP|MAINT & DoNothing (436,475) & DoNothing (165,253) & DoNothing (210,276) & DoNothing (27,556) \\
OP|MAINT|OP|MAINT|OP|DEC & DoNothing (427,470) & DoNothing (161,336) & DoNothing (210,276) & DoNothing (23,325) \\
OP|MAINT|OP|MAINT|OP|MAINT & Repair (518,848) & Repair (199,021) & Repair (314,824) & DoNothing (34,356) \\
\midrule
\multicolumn{5}{c}{\textbf{Failure-regime States}}\\
\midrule
OP|DEC|OP|DEC|OP|DEC & DoNothing (432,155) & DoNothing (176,562) & DoNothing (210,276) & DoNothing (33,010) \\
OP|DEC|OP|DEC|OP|INC & DoNothing (427,470) & DoNothing (169,622) & DoNothing (210,276) & DoNothing (38,801) \\
OP|DEC|OP|DEC|OP|MAINT & DoNothing (443,076) & DoNothing (184,818) & DoNothing (210,276) & DoNothing (23,363) \\
OP|INC|OP|DEC|OP|DEC & DoNothing (438,290) & DoNothing (173,576) & DoNothing (210,276) & DoNothing (23,243) \\
OP|INC|OP|DEC|OP|INC & DoNothing (461,249) & DoNothing (174,372) & DoNothing (210,276) & DoNothing (23,684) \\
OP|INC|OP|DEC|OP|MAINT & DoNothing (434,818) & DoNothing (173,964) & DoNothing (210,276) & DoNothing (23,164) \\
OP|MAINT|OP|DEC|OP|DEC & NA & NA & Repair (314,824) & NA \\
OP|MAINT|OP|DEC|OP|INC & NA & DoNothing (99,198) & Repair (314,824) & NA \\
NOP|INC|OP|MAINT|OP|DEC & DoNothing (274,104) & DoNothing (99,198) & NA & NA \\
NOP|INC|OP|MAINT|OP|INC & DoNothing (274,104) & DoNothing (99,198) & NA & NA \\
NOP|INC|OP|MAINT|OP|MAINT & DoNothing (274,104) & DoNothing (99,198) & NA & NA \\
NOP|DEC|OP|MAINT|OP|DEC & Repair (491,509) & Repair (208,875) & NA & NA \\
NOP|DEC|OP|MAINT|OP|INC & Repair (491,539) & Repair (199,206) & NA & NA \\
NOP|DEC|OP|MAINT|OP|MAINT & Repair (490,808) & Repair (200,314) & NA & NA \\
NOP|DEC|OP|DEC|OP|DEC & Repair (521,318) & Repair (227,412) & NA & NA \\
NOP|DEC|OP|DEC|OP|INC & Repair (521,877) & Repair (242,433) & NA & NA \\
NOP|DEC|OP|DEC|OP|MAINT & Repair (536,573) & Repair (271,407) & NA & NA \\
NOP|DEC|NOP|DEC|OP|DEC & Repair (536,573) & Repair (271,407) & NA & NA \\
NOP|DEC|NOP|DEC|OP|INC & Repair (536,573) & Repair (271,407) & NA & NA \\
NOP|DEC|NOP|DEC|OP|MAINT & Repair (536,573) & Repair (271,407) & NA & NA \\
OP|DEC|NOP|DEC|OP|DEC & NA & NA & DoNothing (210,276) & Repair (45,714) \\
OP|DEC|NOP|DEC|OP|INC & NA & NA & DoNothing (210,276) & Repair (47,364) \\
OP|DEC|NOP|DEC|OP|MAINT & NA & NA & DoNothing (210,276) & Repair (47,364) \\
OP|DEC|NOP|MAINT|OP|DEC & NA & NA & DoNothing (210,276) & NA \\
OP|MAINT|NOP|DEC|OP|DEC & NA & NA & Repair (321,557) & NA \\
OP|MAINT|NOP|DEC|OP|INC & NA & NA & Repair (323,888) & DoNothing (18,654) \\
OP|MAINT|NOP|DEC|OP|MAINT & NA & NA & Repair (323,888) & NA \\
OP|INC|NOP|DEC|OP|DEC & NA & NA & NA & DoNothing (18,654) \\
OP|INC|NOP|DEC|OP|INC & NA & NA & NA & DoNothing (18,654) \\
NOP|MAINT|NOP|DEC|OP|DEC & Repair (517,210) & Repair (269,113) & NA & NA \\
NOP|MAINT|NOP|DEC|OP|INC & Repair (519,177) & Repair (269,346) & NA & NA \\
NOP|MAINT|NOP|DEC|NOP|DEC & Repair (514,603) & Repair (273,523) & NA & NA \\
NOP|MAINT|NOP|MAINT|NOP|DEC & Repair (516,463) & Repair (278,248) & NA & NA \\
NOP|MAINT|NOP|MAINT|NOP|MAINT & Repair (516,463) & Repair (278,248) & NA & NA \\
OP|DEC|NOP|DEC|NOP|DEC & NA & NA & DoNothing (210,276) & Repair (47,364) \\
OP|DEC|NOP|MAINT|NOP|DEC & NA & NA & DoNothing (210,276) & NA \\
OP|DEC|NOP|MAINT|NOP|MAINT & NA & NA & NA & Repair (63,995) \\
OP|INC|NOP|DEC|NOP|DEC & NA & NA & NA & Repair (61,602) \\
OP|INC|NOP|MAINT|NOP|DEC & NA & NA & NA & Repair (63,026) \\
OP|INC|NOP|MAINT|NOP|MAINT & NA & NA & NA & Repair (63,274) \\
OP|MAINT|NOP|DEC|NOP|DEC & NA & NA & Repair (343,105) & DoNothing (18,654) \\
OP|MAINT|NOP|MAINT|NOP|DEC & NA & NA & Repair (349,305) & NA \\
OP|MAINT|NOP|MAINT|NOP|MAINT & NA & NA & Repair (349,327) & NA \\
NOP|DEC|OP|INC|OP|DEC & DoNothing (274,104) & Repair (191,647) & NA & NA \\
NOP|DEC|OP|INC|OP|INC & DoNothing (274,104) & Repair (194,664) & NA & NA \\
NOP|DEC|OP|INC|OP|MAINT & Repair (490,808) & Repair (201,312) & NA & NA \\
\midrule
\multicolumn{5}{c}{\textbf{Recovery-regime States}}\\
\midrule
NOP|INC|NOP|DEC|NOP|DEC & DoNothing (274,104) & DoNothing (99,198) & NA & NA \\
NOP|INC|NOP|DEC|OP|DEC & DoNothing (274,104) & DoNothing (99,198) & NA & NA \\
NOP|INC|NOP|DEC|OP|INC & DoNothing (274,104) & DoNothing (99,198) & NA & NA \\
NOP|INC|NOP|DEC|OP|MAINT & DoNothing (274,104) & DoNothing (99,198) & NA & NA \\
NOP|INC|NOP|MAINT|NOP|DEC & DoNothing (274,104) & DoNothing (99,198) & NA & NA \\
NOP|INC|NOP|MAINT|OP|DEC & DoNothing (274,104) & DoNothing (99,198) & NA & NA \\
NOP|INC|OP|DEC|OP|DEC & DoNothing (274,104) & DoNothing (99,198) & NA & NA \\
NOP|INC|OP|DEC|OP|INC & DoNothing (274,104) & DoNothing (99,198) & NA & NA \\
NOP|INC|OP|DEC|OP|MAINT & DoNothing (274,104) & DoNothing (99,198) & NA & NA \\
NOP|INC|NOP|INC|NOP|DEC & DoNothing (274,104) & DoNothing (99,198) & NA & NA \\
NOP|INC|NOP|INC|NOP|INC & DoNothing (274,104) & DoNothing (99,198) & NA & NA \\
NOP|INC|NOP|INC|NOP|MAINT & DoNothing (274,104) & NA & NA & NA \\
NOP|INC|NOP|INC|OP|DEC & DoNothing (274,104) & DoNothing (99,198) & NA & NA \\
NOP|INC|NOP|INC|OP|INC & DoNothing (274,104) & DoNothing (99,198) & NA & NA \\
NOP|INC|OP|INC|OP|DEC & DoNothing (274,104) & DoNothing (99,198) & NA & NA \\
NOP|INC|OP|INC|OP|INC & DoNothing (274,104) & DoNothing (99,198) & NA & NA \\
NOP|INC|OP|INC|OP|MAINT & DoNothing (274,104) & DoNothing (99,198) & NA & NA \\
OP|INC|NOP|INC|NOP|DEC & DoNothing (274,104) & DoNothing (99,198) & NA & DoNothing (18,654) \\
OP|INC|NOP|INC|NOP|INC & NA & NA & NA & DoNothing (18,654) \\
OP|INC|NOP|INC|OP|DEC & DoNothing (274,104) & DoNothing (99,198) & NA & DoNothing (18,654) \\
OP|INC|NOP|INC|OP|INC & DoNothing (274,104) & DoNothing (99,198) & NA & DoNothing (18,654) \\
OP|INC|NOP|INC|OP|MAINT & DoNothing (274,104) & DoNothing (99,198) & NA & DoNothing (18,654) \\
OP|INC|NOP|MAINT|OP|DEC & NA & NA & NA & DoNothing (18,654) \\
OP|INC|NOP|MAINT|OP|INC & NA & NA & NA & DoNothing (18,654) \\
OP|DEC|NOP|INC|NOP|DEC & NA & NA & DoNothing (210,276) & DoNothing (18,654) \\
OP|DEC|NOP|INC|NOP|INC & NA & NA & DoNothing (210,276) & DoNothing (18,654) \\
OP|DEC|NOP|INC|OP|DEC & NA & NA & DoNothing (210,276) & DoNothing (18,654) \\
OP|DEC|NOP|INC|OP|INC & NA & NA & DoNothing (210,276) & DoNothing (18,654) \\
OP|DEC|NOP|INC|OP|MAINT & NA & NA & DoNothing (210,276) & NA \\
OP|MAINT|NOP|INC|NOP|INC & NA & NA & NA & DoNothing (18,654) \\
OP|MAINT|NOP|INC|OP|INC & NA & NA & NA & DoNothing (18,654) \\
OP|DEC|OP|INC|NOP|INC & NA & NA & DoNothing (210,276) & NA \\
OP|DEC|OP|INC|OP|DEC & DoNothing (424,098) & DoNothing (194,081) & DoNothing (210,276) & DoNothing (23,035) \\
OP|DEC|OP|INC|OP|INC & DoNothing (488,953) & DoNothing (185,383) & DoNothing (210,276) & DoNothing (24,308) \\
OP|DEC|OP|INC|OP|MAINT & DoNothing (439,055) & DoNothing (170,850) & DoNothing (210,276) & DoNothing (25,000) \\
OP|INC|OP|INC|NOP|INC & DoNothing (274,104) & DoNothing (99,198) & NA & DoNothing (18,654) \\
OP|INC|OP|INC|OP|DEC & DoNothing (423,276) & DoNothing (166,988) & DoNothing (210,276) & DoNothing (31,166) \\
OP|INC|OP|INC|OP|INC & DoNothing (412,516) & DoNothing (162,756) & DoNothing (210,276) & DoNothing (36,837) \\
OP|INC|OP|INC|OP|MAINT & DoNothing (431,061) & DoNothing (167,193) & DoNothing (210,276) & DoNothing (26,250) \\
OP|MAINT|OP|INC|OP|DEC & NA & NA & NA & DoNothing (18,654) \\
OP|MAINT|OP|INC|OP|INC & NA & NA & NA & DoNothing (18,654) \\
% \multicolumn{5}{l}{\footnotesize \textit{Note:} OP: operational, NOP: non-operational, INC: increase, DEC: decrease, MAINT: maintain.}
\end{longtable}
\normalsize
\twocolumn

%\begin{figure} 
%    \centering
%    \includegraphics[width=\linewidth]{sensi_repairratio_vs_lambda_p_failed=0.05_pipe23.png}
%    \caption[Repair ratio vs $\lambda$ if the probability of pipe23 failure is $p_{fail}=0.05$]{Repair ratio vs $\lambda$ if the probability of pipe23 failure is $p_{fail}=0.05$.}    \label{fig:repair_ratio_vs_lambda_p_failed=0.05_pipe23}
%\end{figure}

\end{document}